\documentclass[usenatbib]{mn2e}
\usepackage{graphicx,times}

\newcommand{\mgsp}{MGSphot-only }
\newcommand{\mgspns}{MGSphot-only}
\newcommand{\mgss}{MGSphot$+$spec }
\newcommand{\mgssns}{MGSphot$+$spec}
\newcommand{\mpsa}{mag/arcsec$^2$}

\newcommand{\arcdeg}{\mbox{$^\circ$}}

\newdimen\hssize 
\newdimen\hdsize 
\title[Nature and completeness of galaxies detected in 2MASS]{Nature and completeness of galaxies detected in the Two Micron All Sky Survey}

\author[McIntosh et al.]
{Daniel H.\ McIntosh$^{1}$\thanks{E-mail: dmac@hamerkop.astro.umass.edu},
Eric F.\ Bell$^{2}$, Martin D.\ Weinberg$^{1}$, and Neal Katz$^{1}$\\
$^{1}$Department of Astronomy, University of Massachusetts, Amherst, MA 01003, USA\\
$^{2}$Max Planck Institut f\"ur Astronomie, K\"onigstuhl 17, D-69117 Heidelberg, Germany}

\begin{document}

\date{{\sc Draft: } \today }

\pagerange{\pageref{firstpage}--\pageref{lastpage}} \pubyear{2005}

\maketitle

\label{firstpage}

\begin{abstract}
The {\it Two Micron All Sky Survey} (2MASS)
provides the most comprehensive
dataset of near-infrared galaxy properties. We cross correlate
the well-defined and highly complete spectroscopic selection of over
156,000 bright ($r\le17.5$ mag) galaxies in the 
{\it Sloan Digital Sky Survey} (SDSS) Main Galaxy Sample (MGS)
with 2MASS sources to explore the nature and completeness of the
2MASS ($K$-band) selection of nearby galaxies. Using estimates of
total galaxy brightness direct from the 2MASS and SDSS public catalogues,
corrected only for Galactic extinction,
we find that 2MASS detects 90
percent of the MGS brighter than $r=17$ mag. We quantify the completeness 
of 2MASS galaxies in terms of optical properties from SDSS.
For $r\leq16$ mag, 93.1\% of the MGS is found in the 2MASS Extended Source
Catalog (XSC). These detections span the representative range of optical
and near-infrared
galaxy properties, but with a surface brightness-dependent bias 
to preferentially miss sources at the extreme blue and low-concentration
end of parameter space, which are consistent
with the most morphologically late-type galaxy population. An XSC completeness
of 97.5\% is achievable at bright magnitudes, with
blue low-surface-brightness galaxies being
the only major source of incompleteness, if one follows our careful
matching criteria and weeds out spurious SDSS sources.
We conclude that the rapid drop in XSC completeness at $r>16$ mag
reflects the sharp surface-brightness limit of the extended source detection
algorithm in 2MASS. As a result, the $r>16$
galaxies found in the XSC are over-representative in red early types
and under-representative in blue late types. At $r>16$ mag
the XSC suffers an additional selection effect from the $2-3\arcsec$ spatial
resolution limit of 2MASS. Therefore, in the range $16<r\leq17$ mag,
2MASS continues to detect 90\% of the MGS, but with a growing fraction
found in the Point Source Catalog (PSC) only. 
Overall, one third of the MGS is detected in the 2MASS PSC
but not the XSC.
A combined $K\leq13.57$ and $r\leq16$ magnitude-limited
selection provides the most representative inventory of galaxies 
in the local cosmos with near-infrared and optical measurements,
and 90.8\% completeness . Using data from the SDSS Second Data Release,
this sample contains nearly 20,000 galaxies with a median redshift of
0.052.
\end{abstract}

\begin{keywords}
surveys -- galaxies: general -- galaxies: fundamental parameters.
\end{keywords}

\section{Motivation}
\label{sec:intro}

A detailed understanding of the galaxy population in the present-day
universe is essential for constraining theories of galaxy formation
and evolution. To this end, public resources have been dedicated to
producing large comprehensive surveys of galaxies. 
Foremost among these are
the {\it Two Micron All Sky Survey} \citep[2MASS;][]{skrutskie97},
the {\it Sloan Digital Sky Survey} \citep[SDSS;][]{york00}, and
the {\it Two-degree Field Galaxy Redshift Survey} \citep[2dFGRS;][]{colless01},
which will
remain cornerstones of galaxy characterisation for at least the next decade.
2MASS has mapped the near-infrared (near-IR) properties of more than 1.6 million
galaxies over the entire sky, providing a large-scale view of the
local universe minimally affected by dust, and a rich resource for
understanding the full stellar mass component of $z\sim0$ galaxies that is
much less biased by young stellar light compared with optical surveys. 
The SDSS continues to acquire data
and will ultimately cover roughly one quarter
of the entire sky to a limiting depth considerably fainter than 2MASS.
While each survey provides a wealth of measured galaxy properties,
the ability to obtain representative samples of galaxies depends
on a detailed assessment of each survey's selection effects and
completeness. In this paper, we explore the 2MASS selection of local
galaxies using the SDSS Main Galaxy Sample \citep[MGS][]{strauss02}
criteria to define a very complete and well-understood
parent sample for comparison. 

Already the 2MASS Extended Source Catalog \citep[XSC][]{jarrett00a}
has been used to study the large-scale
spatial distribution of galaxies in the local cosmos
\citep{jarrett00b,kochanek03,maller03,maller05,sivakoff05,frith05}.
Additionally, 2MASS has been combined with redshift surveys such as
SDSS, 2dFGRS, and LCRS \citep{shectman96} to quantify the near-IR luminosity and
stellar mass distributions of the present-day galaxy population
\citep{cole01,kochanek01,balogh01,bell03a,bell03b,karachentsev05}.
Our understanding of the
overall $z\sim0$ galaxy population from these studies
depends on the detailed nature of
the 2MASS sample. For example, if
2MASS preferentially
selects red early-type galaxies at all brightnesses, then the
aforementioned results would contain a type-dependent bias. 

In practice, the completeness of a sample is the fraction of sources of
a given magnitude that are detected.
The XSC meets the original 2MASS science requirements of greater than
90\% completeness for the detection of galaxies brighter than
$K<13.5$ mag \citep{jarrett00a}\footnote{This requirement includes
caveats that $K<13.5$ completeness is true only for sources with
Galactic latitude $|b|>20\arcdeg$
and free from stellar confusion. Note, we use $K$ to denote the 2MASS 
$K_{\rm s}$-band (or $K$ short) throughout.}.
The science requirements and initial tests for XSC completeness and
reliability are recorded in the
2MASS Explanatory Supplement\footnote{{\texttt http://www.ipac.caltech.edu/2mass/releases/allsky/doc/explsup.html}}, and 
a useful summary of the findings is given in \S2 of \citet{maller05}.
\citet{cole01} found 2MASS galaxies limited to $K<13.2$ mag
to be highly complete and not missing
a significant fraction of low-surface-brightness systems.
Furthermore, in an initial analysis based
on the SDSS Early Data Release \citep{stoughton02},
we found that the XSC misses only 2.5\% of the known galaxy population
down to $K=13.57$ dust-corrected Kron magnitudes 
\citep{bell03b}. These studies show 2MASS to be a statistically complete
sample of all sources brighter than a fixed $K$-band limit.
Nevertheless, there are reports of galaxy populations
missed by 2MASS; e.g., low-luminosity emission-line
galaxies are under-represented \citep{salzer05}. Also,
Kannappan et al. (in prep) find roughly 40\% of 
Nearby Field Galaxy Survey \citep{jansen00} dwarfs fainter than $M_B=-18$
have no counterparts in the 2MASS XSC.

In this study, we show that the XSC of 2MASS provides a very representative
sample of galaxies with well-defined completeness.
In particular, we use over 156,000 MGS
galaxies from the SDSS Second Data Release
\citep[DR2][]{abazajian04} with spectroscopic redshifts and precise
optical properties to quantify the completeness of 2MASS-selected galaxies.
We elect to use this earlier SDSS release and note that the large sample
size and representative nature afforded by DR2 is more than sufficient for
the analysis herein. Moreover, the SDSS data processing has not changed 
substantively in subsequent releases \citep{adelman05}.
In \S2, we describe the galaxy samples drawn
from SDSS and 2MASS, the relevant parameters we use from these public databases,
and our method for constructing matched galaxy catalogues. We analyse the
completeness of 2MASS-selected galaxies as a function of optical 
properties in \S \ref{sec:gxcompl}. In \S \ref{sec:nature}, we discuss
the nature of galaxies
detected and missed by 2MASS, and we provide our conclusions
in \S \ref{sec:conc}.

\section{Galaxy Samples} 
\label{sec:samples}

For a detailed assessment of the completeness and nature at visible
wavelengths of galaxies found in 2MASS, we require an optical
survey that has enough sky coverage to sample a representative set of
galaxies averaged over all environments, and has a limiting magnitude 
that is deeper than 2MASS so
that we can construct samples of galaxies 2MASS does and does not detect.
The SDSS meets these requirements and includes a suite of photometric
and spectroscopic quantities, including redshifts, for the $>99\%$ complete
and uniform ``main'' spectroscopic selection of galaxies
\citep[MGS;][]{strauss02}.
As such, the MGS is an ideal parent sample to test
the completeness of 2MASS-selected galaxies, quantified
by apparent and intrinsic optical properties.
In this section we outline the relevant aspects of the 2MASS and SDSS
galaxy samples, and our selection of combined samples.
We include galaxies detected as point sources
by 2MASS, and we separate these from extended sources
in the following analyses.

\subsection{SDSS Main Galaxy Sample}

The SDSS is surveying a significant portion of the sky at latitudes outside
of the Galactic plane where the effects of dust on optical
properties are minimal. 
The SDSS provides deep ($r=22.2$ mag limit) photometry from imaging in 
five optical passbands ($ugriz$) and follow-up spectroscopy for the
brighter subset of galaxies.
The target selection algorithm for the MGS is detailed in \citet{strauss02}.
Briefly, useful sources with $>5\sigma$ detection were separated into stars and
galaxies.
The MGS includes all galaxies with $r$-band Petrosian
magnitudes of $r\leq 17.77$, corrected for foreground Galactic extinction
following \citet{schlegel98}.
A further surface brightness limit of $\mu_{50,r}=23.0$ mag/arcsec$^2$
(see \S \ref{sec:optnir} for $\mu_{50,r}$ definition),
retains 99\% of the galaxies with $r\leq 17.77$ mag.
The SDSS galaxy photometry is based on the Petrosian system, which allows
an estimate of the total flux of each galaxy roughly independent
of surface brightness. Recent studies give prescriptions that attempt to
correct these estimates to the actual total \citep{graham05},
but here we elect to leave the SDSS
magnitudes uncorrected. Our analysis is intimately tied to
the MGS selection, which is based on Petrosian magnitudes direct from
the public database, corrected only for Milky Way dust.

Using the SDSS Catalog Archive Server,
we select 319,224 galaxies from the DR2 photometry database 
meeting the MGS magnitude and surface brightness cuts, with
the following additional criteria:
good (status \& 0x2 $>$ 0), primary (mode=1) sources flagged as galaxies
\citep[PrimTarg \& 0x1c0 $>$ 0; i.e., GALAXY or GALAXY\_BIG or
GALAXY\_BRIGHT\_CORE, as defined in Table 27 of][]{stoughton02}
\footnote{The PrimTarg flag is set to GALAXY by the image deblender 
when the MGS criteria are
met including star/galaxy separation, and a number of fiber magnitude cuts
designed to reject bright stars and very bright sources that can contaminate 
adjacent fiber spectra; \citet{strauss02}.}.
The spectroscopic
targeting was designed to meet its completeness specifications at
$r\le17.77$ mag, but in practice this limit
varies\footnote{Spectroscopic targets were selected based on initial processing
of the data resulting in subtle changes in the $r$-band flux limit
and slight variations in the completeness limit over large angular scales.}
between 17.50 and 17.77; therefore, for
statistically representative samples a conservative limit of $r\le17.5$
mag is recommended
\citep{abazajian03}.  We employ this cut resulting in a sample
of 228,645 galaxies with a high degree of completeness. We further remove
all 943 (88) sources with $(g-r)_{\rm mod}>2.5$ ($<-0.5$) colours. Visual
inspection of the SDSS images shows that these colour cuts remove 
defective and contaminated sources. In particular, the red cut sources
are satellite trails and other defects (e.g., false detections)
at $r\le15.5$, and roughly half defects and half point-like contamination at
fainter magnitudes. The blue removed sources are 100\% ($>75\%$) 
point sources at $r\le16.5$ ($16.5<r\le17.5$). We are left
with a magnitude-limited sample of 227,614 galaxies with DR2
photometric properties; we call this our \mgsp sample.

Next, from the DR2 spectroscopic database we select the subset of \mgsp
sources that have useful redshifts\footnote{This value for
the redshift confidence is suggested in the following DR2 website:
{\texttt http://www.sdss.org/dr2/products/spectra/index.html}. We note that
\citet{stoughton02} used a cut of zConf$\geq0.6$ for ``high'' redshift
confidence. The difference between these two cuts amounts to only 10 galaxies.}
(zConf$>0.35$).
This is our primary parent sample of 156,788 galaxies
with photometric properties and reliable redshifts, and represents the
magnitude-limited ($r\leq 17.5$) MGS from DR2; we denote this sample \mgssns.
We stress that the large number difference between the \mgss and \mgsp samples
is the result of the smaller effective coverage of the DR2 spectroscopic
area as compared to the imaging footprint.
A complex tiling algorithm was designed for the spectroscopic survey
\citep{blanton03a} to provide nearly
uniform completeness when sampling the non-uniform large-scale
galaxy distribution of the local universe. For SDSS, this tiling samples
$92-93\%$ of all targets, with the incomplete coverage owing to multiple
objects within the $55\arcsec$ minimum fiber separation of the
SDSS multi-object spectrograph (i.e., ``fiber collisions''). 
This leads to a slight
systematic under representation in regions of high galaxy number density
\citep{hogg04}. For our purpose we are concerned with the completeness
and representative nature of the overall sample, and therefore, a
detailed accounting of MGS completeness as a function of position
is not required.

Overall, the SDSS spectroscopic survey is 90\% complete
\citep{blanton03d,hogg04}, with 7\% incompleteness owing to fiber
collisions as described above, and additional sources of incompleteness
from bright star contamination (2\%) and spectra with incorrect
or impossible to determine redshifts (1\%). The effective sky
coverage of the DR2 spectroscopy is 2627 square degrees \citep{abazajian04}.
If we extend the magnitude cut to $r=17.77$, we obtain 209,730 MGS
sources with redshifts, or 79.84 deg$^{-2}$. This is in good agreement
with the expected surface density of $\approx80$ deg$^{-2}$ from the MGS
targeting of $\approx90$ deg$^{-2}$ with a completeness of 90\%, less an
additional 1\% from the $\mu_{50,r}>23.0$ mag/arcsec$^2$ cut.

Finally, we note that 
two effects cause the MGS spectroscopic sample to become noticeably
incomplete for galaxies brighter than $r=14.5$ mag: 
(i) to avoid excessive cross-talk and/or saturation in the spectrograph
detectors, some targets were rejected if any pixels exceeded a peak
flux level; and (ii) the image deblending software sometimes
over deblended galaxies with large angular extent \citep{strauss02}.
We performed a
preliminary examination of this issue by using 2MASS to identify the
locations of all large bright galaxies missed by the pipeline processing
in a 710 deg$^2$ region (54\% of spectroscopic coverage) of the 
SDSS First Data Release \citep{abazajian03}. We found that the MGS
had no detection for $>20\%$ and $>50\%$ of XSC sources brighter than $K=11.5$ 
and $K=9.5$ mag, respectively. From a typical colour of $(r-K)\approx3$ 
\citep[e.g.,][]{fukugita95}, these $K$ limits correspond roughly to
$r=14.5$ and $r=12.5$ mag. Yet, the MGS
did not preferentially miss or detect galaxies of one morphological type
(visually early or late) over another. Therefore, the bright end
of the MGS remains representative. Furthermore, we note that an 
$r\leq14.5$ mag cut contains 1.5\% of the total $r\leq17.5$ MGS;
therefore, bright galaxy incompleteness only affects our overall
completeness at the $<1\%$ level.
Throughout this work, our determinations of 2MASS completeness are
with respect to the MGS, limited only at the faint end ($r=17.5$).
We will address this issue in more detail in a subsequent paper.

\subsection{2MASS Extended and Point Source Catalogs}

2MASS is an all-sky survey with uniform photometry in the near-IR
$JHK$ ($1.15,1.65,2.15\micron$) passbands.  The 2MASS analysis pipeline
uses well-defined selection criteria to separate extended and point
sources. The pipeline and XSC construction are described in detail
in \citet{jarrett00a}. The photometric calibration and uniformity
of 2MASS is quantified in \citet{nikolaev00}.

The PSC provides magnitudes for 188 million sources brighter than $K=14.3$ mag, the
$10\sigma$ point-source sensitivity limit. At this magnitude limit, 
the PSC meets the 2MASS science requirement of 99\% completeness (see the
Explanatory Supplement). PSC magnitudes given in
$8\arcsec$ diameter apertures (k\_m\_stdap)
provide a rough total brightness
for unresolved galaxies. In \citet{bell03b}, we showed that
PSC aperture magnitudes for galaxies fainter than $K=13.5$ have rather large
systematic offsets of -0.85 mag (0.3 mag rms). 
We do not recommend using the PSC for precise
photometric studies of galaxies, but
we include PSC detections in this
analysis to provide a complete picture of 2MASS galaxy selection.

The XSC contains over 750,000 galaxies brighter than $K=13.5$ mag.
As described in \S1, at this magnitude limit the
XSC meets the 2MASS science requirement of at least 90\% galaxy completeness
for the Galactic latitudes ($|b|>20\arcdeg$) covered by SDSS.
In addition, within these limits the XSC is at least 98\% reliable; i.e.,
it contains true extended astronomical objects.
The automated processing of the XSC produces systematically
incomplete photometry for the small subset of galaxies larger 
than $50\arcsec$ owing to the
typical 2MASS scan width of $8.5\arcmin$ \citep{jarrett00a}. 
For our analysis, we include
the 2MASS Large Galaxy Atlas \citep{jarrett03} sample of
540 galaxies flagged (cc\_flag=`Z') in the XSC database, which was
assembled to account for most of the scan size photometric incompleteness.

2MASS employs the \citet{kron80} system to define total galaxy 
magnitude in a fiducial elliptical aperture
with semi-major axes equal to 2.5
times the first moment of the brightness distribution.
The first moment calculation is computed to a radius that is
five times the $J$-band 20 mag/arcsec$^2$ isophote and is
limited to a minimum of $5\arcsec$ because the point-spread function
(PSF) dominates
the surface brightness profile within this radius (see the
2MASS Explanatory Supplement for details). Throughout this paper
all 2MASS photometry will be cited in extinction-corrected Kron magnitudes.
As a result of the short exposure times (7.8 s with a 1.3-m telescope),
2MASS Kron magnitudes systematically underestimate
the true total flux by $\approx0.1$ mag \citep{cole01,bell03b}. The purpose
of this work is to describe the galaxy content of the 
publicly-available 2MASS data; thus, throughout we do not correct 
apparent near-IR magnitudes for the missed flux. We note that we do
consider this correction when estimating $k$-corrections
(see \S \ref{sec:optnir}).

\subsection{Optical and Near-IR Quantities}
\label{sec:optnir}

Besides the Petrosian and Kron measures of total apparent magnitude
discussed previously,
we use additional optical and near-IR quantities in our analysis.
We quantify galaxy
colour using $(J-K)$ Kron magnitudes and $(g-r)$ model magnitudes.
The SDSS includes model magnitudes for all five passbands measured
consistently with the same aperture, which is
determined by the best-fit model (either a de Vaucouleurs or an exponential
radial profile) to the $r$-band image with PSF convolution. Colours based on
these model magnitudes are recommended for galaxy studies \citep{abazajian04}.

The spectroscopic redshifts from the SDSS allow us to calculate rest-frame
total luminosities and colours, $k$-corrected to redshift zero.  We assume $H_0 =
70$\,km\,s$^{-1}$\,Mpc$^{-1}$, $\Omega_M=0.3$, and $\Omega_{\Lambda}=0.7$.
We estimate $k$-corrections using the method described by \citet{bell03b}.
Briefly, we fit a
grid of non-evolving stellar population models 
\citep[{\sc P\'egase}][]{fioc97}, with a range of metallicities
and star formation histories of the form $\Psi \propto \exp{(-t/\tau)}$,
to the optical and near-IR photometry of each galaxy and derive the
$k$-correction from the best-fit template. Here,
$\Psi$ is the star formation rate and $\tau$ is the $e$-folding
timescale, and the star formation starts at $z_{\rm initial}=4$.
For the optical
photometry, we adopt Petrosian $gri$ magnitudes and use model colours to
estimate higher signal-to-noise $u$ and $z$ magnitudes: 
$u^{\prime}=r+(u-r)_{\rm mod}$ and $z^{\prime}=r+(z-r)_{\rm mod}$.
When near-IR photometry is available (i.e., for galaxies with 2MASS
detections), we apply a uniform offset of $-0.1$ magnitude to $J$ and $K$
to bring the 2MASS Kron magnitudes into
agreement with magnitudes from deeper $J$ and $K$-band data, following
\citet{cole01} and \citet{bell03b}. 
As already stated, we correct all magnitudes, luminosities, and colours
for foreground reddening using the Galactic dust map of \citet{schlegel98}. 
The mean and rms extinction corrections for the data we
analyse here are $-0.099\pm0.071$ ($r$-band), $-0.013\pm0.009$ ($K$-band),
$-0.038\pm0.027$ ($g-r$), and $-0.019\pm0.014$ ($J-K$) mag.

We include further galaxy structural properties in this analysis.
The SDSS and 2MASS databases each provide measures of galaxy orientation
given by the axis ratio ($b/a$): 2MASS measures $(b/a)_K$ from an elliptical
fit to the $3\sigma$ isophote in the $K$-band image, 
whereas from SDSS we take the average $r$-band axis ratio
$\left<b/a\right>_r$ from values derived from the pure 
de Vaucouleurs and pure exponential model fits to the light profile.
SDSS provides circular radii ${\rm r}_{90,r}$ and ${\rm r}_{50,r}$ 
that contain 90\% and 50\% of the Petrosian flux, respectively.
The 2MASS XSC contains elliptical half-light sizes ${\rm a}_{50,K}$ measured 
along the major axis; therefore, for consistency we calculate the circularised
$K$-band radius ${\rm r}_{50,K}={\rm a}_{50,K}\sqrt{(b/a)_K}$.
In turn, the sizes can be used to calculate various quantities such
as the concentration index $c_r={\rm r}_{90,r}/{\rm r}_{50,r}$, which
is a measure of the degree of central concentration in the $r$-band galaxian
light profile. In addition, within the half-light radii we have
average surface brightnesses 
$\mu_{50,r}= r + 2.5\log_{10}(2\pi{\rm r}_{50,r}^2)$ and
$\mu_{50,K}= K + 2.5\log_{10}(2\pi{\rm r}_{50,K}^2)$.
We calculate physical sizes at each redshift from the angular diameter
distance and our assumed cosmological parameters.

\subsection{Combining Galaxy Detections From 2MASS and SDSS}
\label{sec:xcorrl}

We cross correlate the positions of our MGS-based samples with
the XSC and PSC from 2MASS. For matching to the 2MASS XSC, we employ an 
optimum set of criteria (see Appendix \ref{app:xcorrl})
that minimises contamination from multiple and random matches.
We look for PSC matches only for those remaining MGS galaxies not detected 
in the XSC using
a simple criteria of closest match within a $2\arcsec$ radius.
The mean and rms angular separations between 2MASS and SDSS galaxy positions
are $0.45\arcsec$ and $0.22\arcsec$ for the XSC, and $0.35\arcsec$ and
$0.29\arcsec$ for the PSC. 
The absolute astrometric accuracies are $0\farcs5$ rms for 2MASS
\citep{jarrett00a} and $<0\farcs1$ rms for SDSS with $r<20$ mag
\citep{pier03}. 

This cross correlation provides three subsamples of our main parent
sample of 156,788 \mgss galaxies with DR2 spectroscopic measurements:
(1) 89,742 galaxies with 2MASS extended source matches (and therefore
have $ugrizJK$ total galaxy fluxes, plus optical and near-IR properties),
(2) 51,716 with 2MASS point source counterparts (thus have $ugrizJK$ fluxes
and optical properties), and
(3) 15,330 without any 2MASS source matches (thus optical data only).
Likewise, we find the following subsets from the cross correlation of
the \mgsp sample ($N=227,614$ galaxies) with 2MASS:
(4) 130,348 matched to XSC, (5) 73,335 matched to PSC, and
(6) 23,931 without any 2MASS counterparts.
A further 2MASS completeness cut of $K\le13.57$ to subsample (1)
results in 43,663 galaxies -- the largest complete
sample of galaxies with optical and near-IR photometric
properties and reliable redshift measurements to date.
We plot the redshift distributions of
2MASS extended sources with spectroscopic identification from DR2 in
the upper right panel of Figure \ref{fig:all_cfs}. 
The median redshift of the $K\le13.57$ sample
is $z_{\rm med}=0.076$, compared with $z_{\rm med}=0.104$ 
for the MGS to $r\leq17.77$ mag
\citep{strauss02}; we find the MGS cut at $r=17.5$ mag has $z_{\rm med}=0.092$.

\begin{figure*}  
\center{\includegraphics[scale=0.85, angle=0]{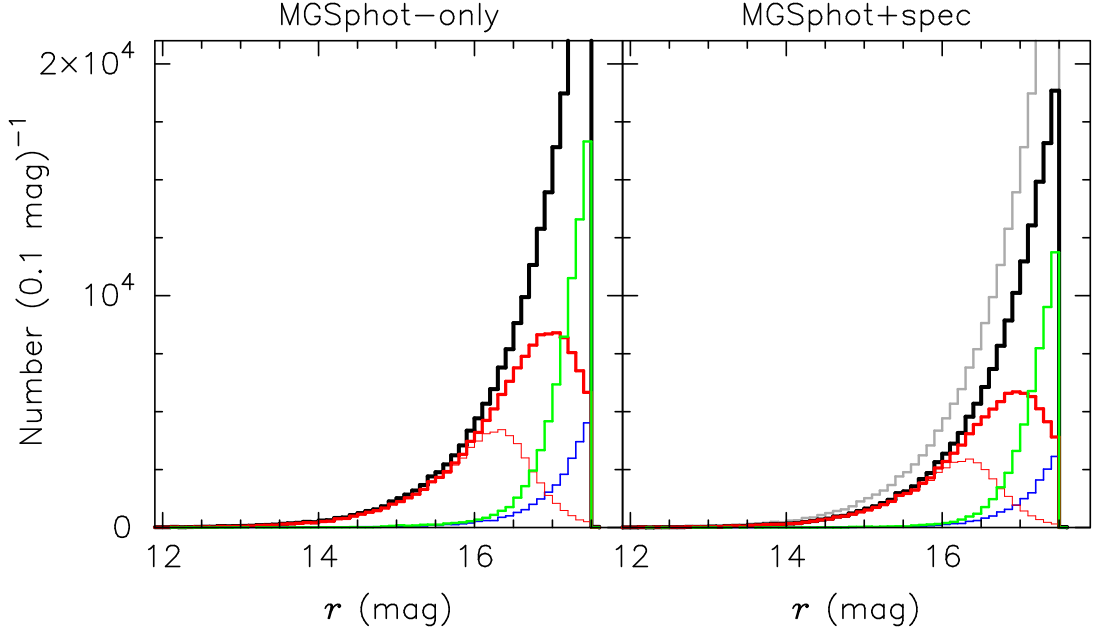}}
\caption{Apparent $r$-band Petrosian magnitude distributions (black) 
for the MGSphot-only ($N=227,614$) and
MGSphot$+$spec ($N=156,788$) primary samples selected from DR2
and limited to $r \le 17.5$ mag.  
In each panel, we show the related subsamples from our cross
correlation with the 2MASS XSC (red) and PSC (green); sources without 2MASS
counterparts are shown in blue. For comparison, we plot
the MGSphot-only distribution
in grey in the right panel. Thin red lines denote 
$K \le 13.57$ magnitude-limited XSC samples.
All magnitudes are corrected for Galactic extinction following
\citet{schlegel98}.
\label{fig:rhist}}
\end{figure*}

\begin{figure}  
\center{\includegraphics[scale=0.85, angle=0]{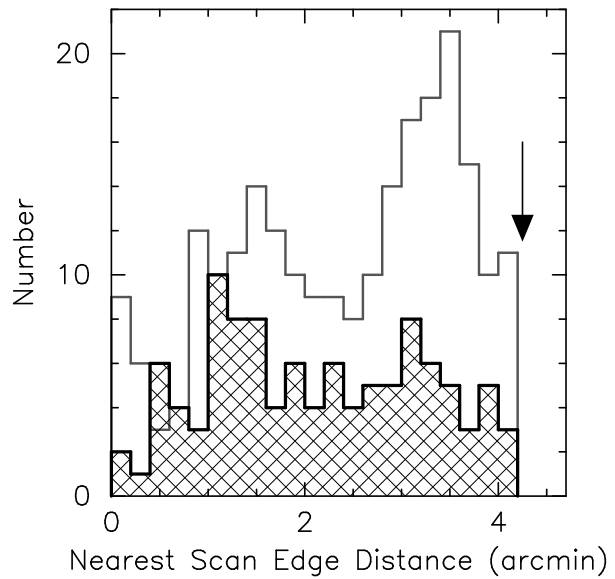}}
\caption{Distribution of distances from the nearest image scan edge for
bright \mgss sources not detected by 2MASS in either the XSC or PSC.
Each 2MASS scan is $8.5\arcmin \times 6\arcdeg$, thus, the maximum
distance to the nearest edge is $4.25\arcmin$ as shown by the arrow.
We plot all $r\leq15$ (hatched), and a random half of the $15<r\leq16$ 
(dark grey outline), nondetections separately.
Bright sources missed by 2MASS are uncorrelated with the scan boundaries.
\label{fig:edgedist}}
\end{figure}

With the set of fundamental optical properties provided by SDSS,
we compare the \mgsp and \mgss parameter distributions
in Table \ref{tab:phot_spec_comp}. In terms of mean, RMS, and median of
each optical parameter distribution, the main photometric sample
and the subset with spectroscopy appear indistinguishable. In 
Figure \ref{fig:rhist}, we plot the $r$ magnitude distributions for
the \mgspns, \mgssns, and corresponding subsamples.
In general, we see that the XSC detects most MGS galaxies brighter
than $r=16$ mag, that much of the fainter MGS sources are
detected in the 2MASS PSC, and that the galaxies not detected by 2MASS
are typically faint ($r>16$ mag). We note that MGS nondetections brighter
than $r=16$ are unrelated to the 2MASS scanning geometry
(see Fig. \ref{fig:edgedist}).
Nevertheless, Figure \ref{fig:rhist} is
not very illuminating in regards to the details of which galaxies 2MASS
detects as extended or point sources, or not all.

\begin{figure*}  
\center{\includegraphics[scale=0.85, angle=0]{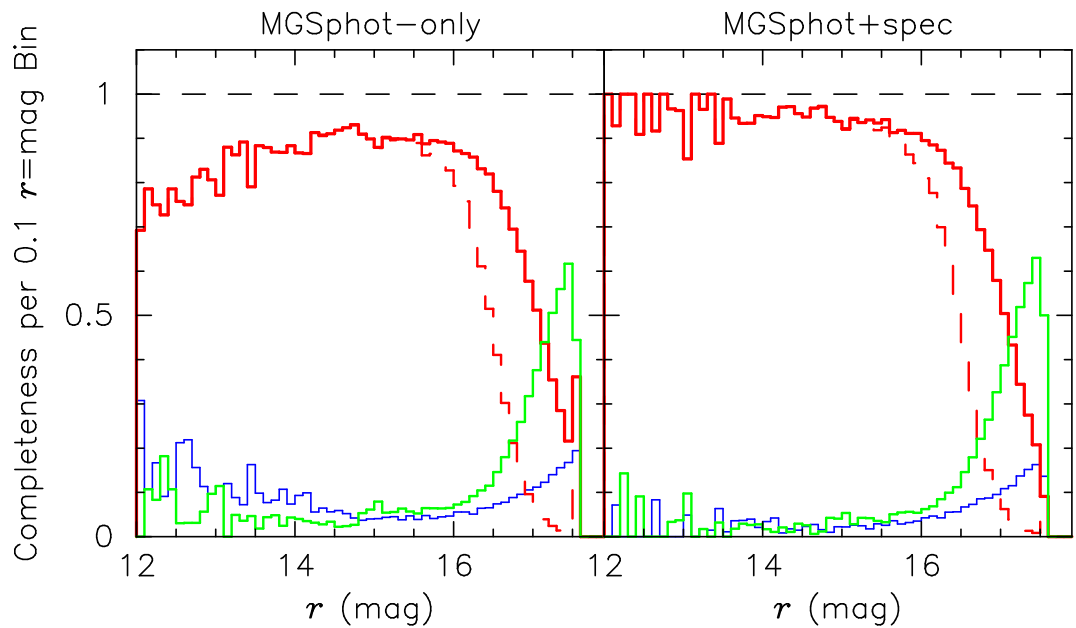}}
\caption{Completeness of subsamples relative to the
\mgsp (left) and \mgss (right)
primary samples as a function of extinction-corrected
$r$-band magnitude. For each $\Delta r=0.1$
mag bin, we plot the fraction of MGS sources with XSC (red), PSC (green),
and no (blue) matches relative the
normalised primary samples (horizontal dashed lines). We plot
the $K \le 13.57$ mag XSC matches as dashed red lines. In the left panel
we note that the XSC appears to be badly incomplete at magnitudes brighter
than $r=15$. We stress that this is the result
of contamination by bright image artifacts in the \mgspns, rather than
actual bright sources missed by the XSC (see \S
\ref{sec:xcorrl}). The right panel shows the actual XSC completeness,
which is $>90\%$ for $r\leq16$ mag.
\label{fig:rmag_cf}}
\end{figure*}

In Figure \ref{fig:rmag_cf},
we show the $r$-magnitude ``completeness function'' (CF) for each
subsample of \mgsp and \mgssns. For example, the XSC (red) $r$-magnitude CF is
given by the number of XSC-MGS matches divided
by the total number of MGS sources in bins of $\Delta r=0.1$ mag.
We further quantify the
completeness of 2MASS-detected galaxies in \S \ref{sec:compl} by analysing the
CFs relative to the MGS using additional optical properties.
Relative to \mgspns, the 2MASS XSC
appears $\ge10\%$ incomplete at all $r$ magnitudes, with systematically
increasing incompleteness with increasing brightness. Yet, when considering
spectroscopically-confirmed galaxies, the $r$-mag CF of the XSC shows
that 2MASS galaxy detections are $>90\%$ complete in every magnitude bin
down to $r=16$. If we
include the galaxies detected in the PSC, the overall galaxy detection
completeness of 2MASS using the \mgss sample remains 
$>90\%$ for all $r\leq17$ mag bins, and $>80\%$ for bins in the range
$17.0<r\leq17.5$ mag.

\begin{figure}  
\center{\includegraphics[scale=0.85, angle=0]{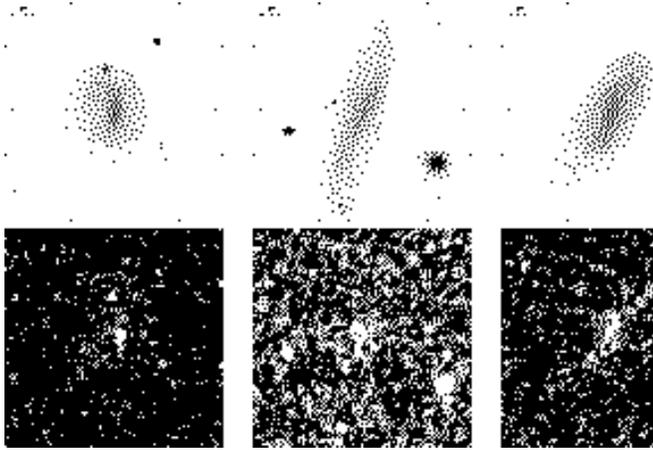}}
\caption{Examples of three bright ($r\leq14.5$ mag) LSBs
from \mgsp without matches in 2MASS. From left to right,
the J2000.0 coordinates are (09:08:51.02,+03:26:54.0), 
(15:04:30.15,-00:51:04.8) and (12:11:19.92,+01:29:32.2). For each case, we show
the colour image from SDSS (top) and the 2MASS $K$-band image from the
same region of sky (bottom). These large, bright galaxies are too
low-surface-brightness to be detected by the 2MASS pipeline.
Images are $101\arcsec$ on a side, east is to the left, and a $10\arcsec$
line is given in the upper left of each optical image from SDSS.
\label{fig:noLSB}}
\end{figure}

The incompleteness of 2MASS in the \mgsp
sample at bright $r$ magnitudes appears to be from sources not
detected in 2MASS in either the XSC or the PSC. This is surprising
given that the 2MASS XSC is reported to be 97.5\% complete for bright 
($K\le13.57$ mag) sources \citep{bell03b}. To ascertain the nature of this
apparent incompleteness, we visually inspect SDSS images
of the 346 unmatched
\mgsp sources with $r\le14.5$ mag and find that only a small fraction (13\%)
are actual bright galaxies. The vast majority (193) of these unmatched, bright
SDSS photometric sources are imaging artifacts from luminous stars (e.g., 
diffraction spikes, scattered light). An additional 109 are false detections,
the result of scattered halo light from nearby large galaxies or the 
overdeblending of large galaxies. The small number
of $r\le14.5$ SDSS galaxies without 2MASS counterparts have large angular
sizes and thus have very low apparent surface brightness. 
As shown in Figure \ref{fig:noLSB}, typical examples of these 
low-surface-brightness galaxies (LSBs)
appear in 2MASS $K$-band images but are below the detection
pipeline threshold. We note that if we correct the \mgsp $r$ mag distribution
above $r=14.5$ mag for the 302 artifacts and false detections, then the
fraction of bright SDSS galaxies with XSC matches (3421/3621)
agrees with that found for the \mgss XSC subsample (95\%).

\begin{figure*}  
\center{\includegraphics[scale=0.8, angle=0]{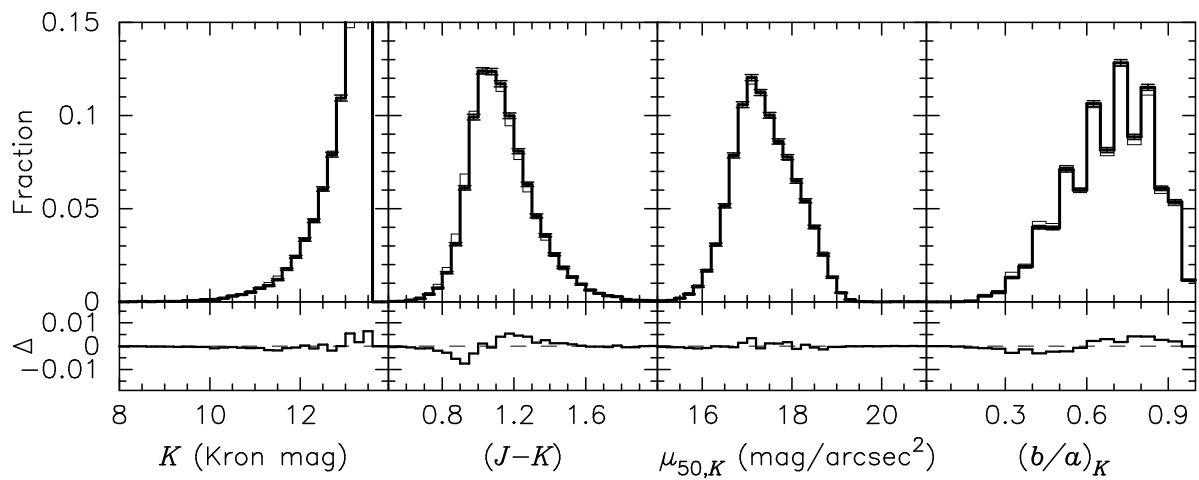}}
\caption{Comparison of two selections of 2MASS galaxies with $K\leq13.57$ mag
(97.5\% completeness): (1) the 548,132 XSC objects
outside of the Galactic plane ($|b|>20\arcdeg$) with galaxy-like
colours of $0.1\le(J-K)\le2.5$ (thin line), and (2) the 43,663 XSC galaxies 
matched to the $r\le17.5$ DR2 spectroscopic sample (bold line).
In each panel we compare the DR2 region and
overall high-latitude relative distributions for four primary parameters:
(from left to right) $K$-band Kron magnitude, $(J-K)$ colour,
$K$-band mean surface brightness within ${\rm r}_{50,K}$, 
and $K$-band axis ratio. 
For each parameter, the difference (DR2$-$ high-latitude)
between the relative fractions is given at the bottom of
each panel.  We show the Poisson
errors per bin for the 2MASS-SDSS matched subsample distributions,
which are smaller typically than the per-bin differences.
The subsample of $K_c\leq13.57$ 2MASS galaxies with DR2 counterparts
is representative of the full 2MASS XSC outside of the Galactic plane.
\label{fig:rep4} }
\end{figure*}

For the rest of this analysis we will use the \mgss ($\equiv$ MGS)
as the parent sample for describing the completeness of 2MASS-selected
galaxies. This choice circumvents the artifact contamination shown above
for \mgspns, and provides redshift information 
and spectroscopic confirmation for each galaxy.
It is important to demonstrate the representative nature of 2MASS galaxies
within the DR2 region and the $K\leq13.57$ mag limit, where the XSC is
required to be 90\% complete and 98\% reliable. 
In Figure \ref{fig:rep4}, we
compare the 2MASS near-IR property distributions for 43,663 XSC galaxies 
that have \mgss counterparts
and 548,132 objects from the XSC with $|b|>20\arcdeg$ and
$0.1\le(J-K)\le2.5$. Note, the additional colour cut
excludes 201 sources that are likely to be high-latitude
Milky Way extended emission.
We do not compare galaxies at fainter magnitudes, where
2MASS becomes increasingly nonuniform, unreliable, and incomplete.
We see that the $K$-magnitude-limited samples have nearly identical 
$K$, $(J-K)$, $\mu_{50,K}$, and $(b/a)_K$ relative distributions.
There is a small $(J-K)$ difference such that
slightly more red galaxies are found in the MGS region 
than the overall sky above
and below the Galactic plane. Following the same procedure as in
\citet{bell03b}, we find that the density of $10.0<K<13.5$ XSC sources in
the DR2 spectroscopic coverage area is $n_{\rm gal}=17.94 {\rm deg}^{-2}$,
or 1.1\% larger than the sky outside of the Milky Way plane. We expect
that the slight red galaxy excess results from the small overdensity of DR2.
The disagreement of completeness
in XSC matches to bright MGS sources seen in Figure \ref{fig:rmag_cf}
is the result of differences in the \mgsp and \mgss selection rather
than that of 2MASS.

\begin{table}
\caption{Comparison of \mgsp and \mgss parameter distributions.}
\label{tab:phot_spec_comp}
\begin{tabular}{lcccccc}
\hline\hline
  & \multicolumn{3}{c}{MGSphot-only} & \multicolumn{3}{c}{MGSphot$+$spec} \\
\cline{2-4} \cline{5-7} \\
Parameter & mean & rms & median & mean & rms & median \\
\hline
$r$ (mag)           & 16.73  & 0.74  & 16.96  & 16.75  & 0.72  & 16.97\\
$c_r$               & 2.64  & 0.42  & 2.65   & 2.64  & 0.41  & 2.65\\
$(g-r)$ (mag)       & 0.81   & 0.26  & 0.84   & 0.81   & 0.25  & 0.84\\
$\mu_{50,r}$ (mag/arcsec$^2$) & 20.75  & 0.83  & 20.76  & 20.76  & 0.78  & 20.77\\
$r_{50,r}$ (arcsec) & 2.82   & 1.69  & 2.46   & 2.77   & 1.40  & 2.46\\
$(b/a)_r$           & 0.63   & 0.21  & 0.67   & 0.64   & 0.21  & 0.67\\
$z$                 &        &       &        & 0.097  & 0.051 & 0.092\\
\hline
\end{tabular}
\vskip 8pt
\begin{minipage}{\hssize}
Characteristics (mean, RMS, and median) of the parameter
distributions from the MGSphot-only ($N=227,614$ galaxies) and
the MGSphot$+$spec ($N=156,788$ galaxies) samples.
\end{minipage}
\end{table}

\section{2MASS Galaxy Completeness}
\label{sec:gxcompl}

We now use the SDSS main spectroscopic sample to
explore the completeness of 2MASS galaxies in more detail.
Recall that from the 156,788 $r\leq17.5$ mag galaxies from the \mgssns,
we match 89,742 in the XSC and 51,716 in the PSC, leaving 15,330 without
2MASS counterparts.
In Figure \ref{fig:all_cfs}, we quantify the completeness of 2MASS-selected
galaxies as a function of six optical observables: (from left to right)
$r$-band concentration $c_r$, $(g-r)$ model colour, average half-light
surface brightness $\mu_{50,r}$, half-light radius ${\rm r}_{50,r}$, 
mean axis ratio $\left<b/a\right>_r$, and spectroscopic redshift $z$. 
In the upper row
we plot the total distributions for each property. The black histogram
shows the \mgss sample, the 2MASS matches are in red (XSC) and green (PSC),
and the blue histogram plots SDSS galaxies not detected by 2MASS.
In the lower rows of Figure \ref{fig:all_cfs},
we give the 2MASS galaxy CFs for each property split into
four $r$ magnitude bins each with the following minimum XSC
completeness per bin from inspection of Figure \ref{fig:rmag_cf}:
95\% ($r\le15.0$), 90\% ($15.0<r\le16.0$), 50\% ($16.0<r\le17.0$),
and 10\% ($17.0<r\le17.5$).
For this analysis the most important completeness pertains to galaxies
detected in the XSC (thick red line), which provides a set of well-defined 
galactic parameters at near-IR wavelengths. We also show the CFs for the
PSC (green) and unmatched (blue) galaxies in each $r$ mag bin.
The CFs for each subsample demonstrate and quantify the selection effects
of galaxy detection by 2MASS; therefore, it is useful to think of the
CFs as selection functions.
The thin red line represents the CF for XSC matches brighter than
$K=13.57$ mag. Here we plot the CFs with variable binning
such that the number of \mgss sources 
per parameter interval is constant.
For comparison, we show the \mgss property distributions
(grey histograms) for each $r$ magnitude interval, and
we tabulate the breakdown of galaxies within these bins in
Table \ref{tab:rbin_totals}.

\begin{table}
\caption{Galaxy counts per $r$-magnitude interval.}
\label{tab:rbin_totals}
\begin{tabular}{cccccc}
\hline\hline
 Interval & \multicolumn{5}{c}{Number}\\
\cline{2-6}\\ 
 & Total & XSC & XSC cut & PSC & Unmatche \\
\hline
$r\le15.0$ &       5012 &  4780 &  4777 &   124 &   108\\
$15.0<r\le16.0$ & 16093 & 14864 & 14379 &   759 &   470\\
$16.0<r\le17.0$ & 60807 & 45061 & 22567 & 11379 &  4367\\
$17.0<r\le17.5$ & 74876 & 25037 &  1940 & 39454 & 10385\\
Totals          & 156788 & 89742 & 43663 & 51716 & 15330\\
\hline
\end{tabular}
\vskip 8pt
\begin{minipage}{\hssize}
For each extinction-corrected
Petrosian $r$-magnitude interval (1), we give the total number of galaxies (2)
from the MGS spectroscopic sample (MGSphot+only) and the number of matches
found in the XSC (3), the XSC with a faint $K=13.57$ mag cut (4), the PSC (5),
and the remaining amount that have no matches in either the XSC or PSC (6).
These magnitude intervals correspond to average XSC galaxy completeness
values of 95.4, 92.4, 74.1, and 33.4 percent. 
\end{minipage}
\end{table}

\subsection{Conditions for Which the XSC is $>90$ Percent Complete}
\label{sec:compl}

At a magnitude cut of $r\le15$ (Fig. \ref{fig:all_cfs}, second row of panels),
2MASS detects galaxies in the XSC at
better than 95\% completeness. The small fraction not
detected in the XSC is from blue ($g-r<0.5$), low concentration ($c_r<2$)
systems
at low redshift ($z<0.02$), split roughly between galaxies not in 2MASS
at all (i.e., not in the XSC or PSC), and galaxies detected by 2MASS as 
point-sources. The nondetections have large angular sizes 
(${\rm r}_{50,r}>10\farcs$) and, hence, low surface brightnesses
($\mu_{50,r}>22$~\mpsa). The remaining galaxies missed by the XSC
have high-surface-brightness ($\mu_{50,r}<18.5$~\mpsa) and are
detected in the PSC, yet with systematically elongated axis ratios
($\left<b/a\right>_r<0.3$). As expected, the galaxies in the PSC are
unresolved by 2MASS with ${\rm r}_{50,r}<3\arcsec$; therefore, the trend
towards an over representation of $\left<b/a\right>_r<0.3$ galaxies in the PSC
at all $r$ magnitude intervals
is the result of different spatial resolutions for SDSS
\citep[$1\farcs4$ median][]{abazajian04} and 2MASS ($2-3\arcsec$)
\citep{jarrett03}.
At fainter $r$, the PSC axis ratio CFs are nearly flat, which
means that galaxies detected by the PSC are
not preferentially round, they are merely un-resolved.
In general,
apparently bright galaxies that 2MASS does not detect
in the XSC are concentrated at the extremes of
the property distributions observed at optical wavelengths. In other words,
2MASS detects $r\le15$ mag galaxies at 95.4
broad range of normal optical properties. 

We provide an alternative way to view the completeness of the XSC
in Figure \ref{fig:rmagsize}. 
In terms of absolute quantities, 
we show the $r$-band luminosity-size plane
($M_r-{\rm r}_{50,r}$; i.e., Petrosian absolute magnitude versus half-light
radius) and we see that  for $r\leq15$ the 108 nondetections 
(blue circles)
are intrinsically small and faint in visible red light.
In contrast, we find that the 124 PSC-only detections (green triangles) occur
throughout the same general envelope of absolute magnitude
and size as the the nearly complete XSC sample shown by the red contours.

The XSC galaxy completeness remains $>90\%$ per $r$ mag bin, and 92.4
average, for $15.0<r\le16.0$
mag (third row in Fig. \ref{fig:all_cfs}). 
The incompleteness continues to be dominated by nearby sources with $z<0.04$,
of which 60\% are detected in the PSC. The rest are too low-surface-brightness
to be detected by 2MASS in the XSC or PSC. 
The XSC is most complete for concentrated $c_r>2.6$, red
($g-r>0.6$) galaxies with $\mu_{50,r}<21.5$~mag/arcsec$^2$.
We note that for $r\le16.0$ mag, there is little difference in the completeness
for all sources from the XSC, and the subset brighter than
$K=13.57$ mag (thin red lines).
At the resolution limit of 2MASS, about half of the
${\rm r}_{50,r}<2\arcsec$ MGS galaxies,
some of which have red colours, are found in the PSC; and the rest have
XSC detections. As seen at brighter $r$ magnitudes, the known galaxies 
that altogether escape 2MASS detection tend to be blue LSBs with low
concentration indices. As such, we see these galaxies are confined to
a different part of the
$M_r-{\rm r}_{50,r}$ plane than the bulk of normal galaxies detected
by the XSC (see upper right panel of Fig. \ref{fig:rmagsize}). This
behaviour for the nondetections is precisely what is expected in a
surface brightness-limited survey. We note that as with the brighter
cases, the un-resolved
galaxies found in the PSC continue to span the full range of $M_r$ and intrinsic
${\rm r}_{50,r}$ populated by XSC matches (red contours).

\begin{figure*}  
\center{\includegraphics[scale=0.85, angle=0]{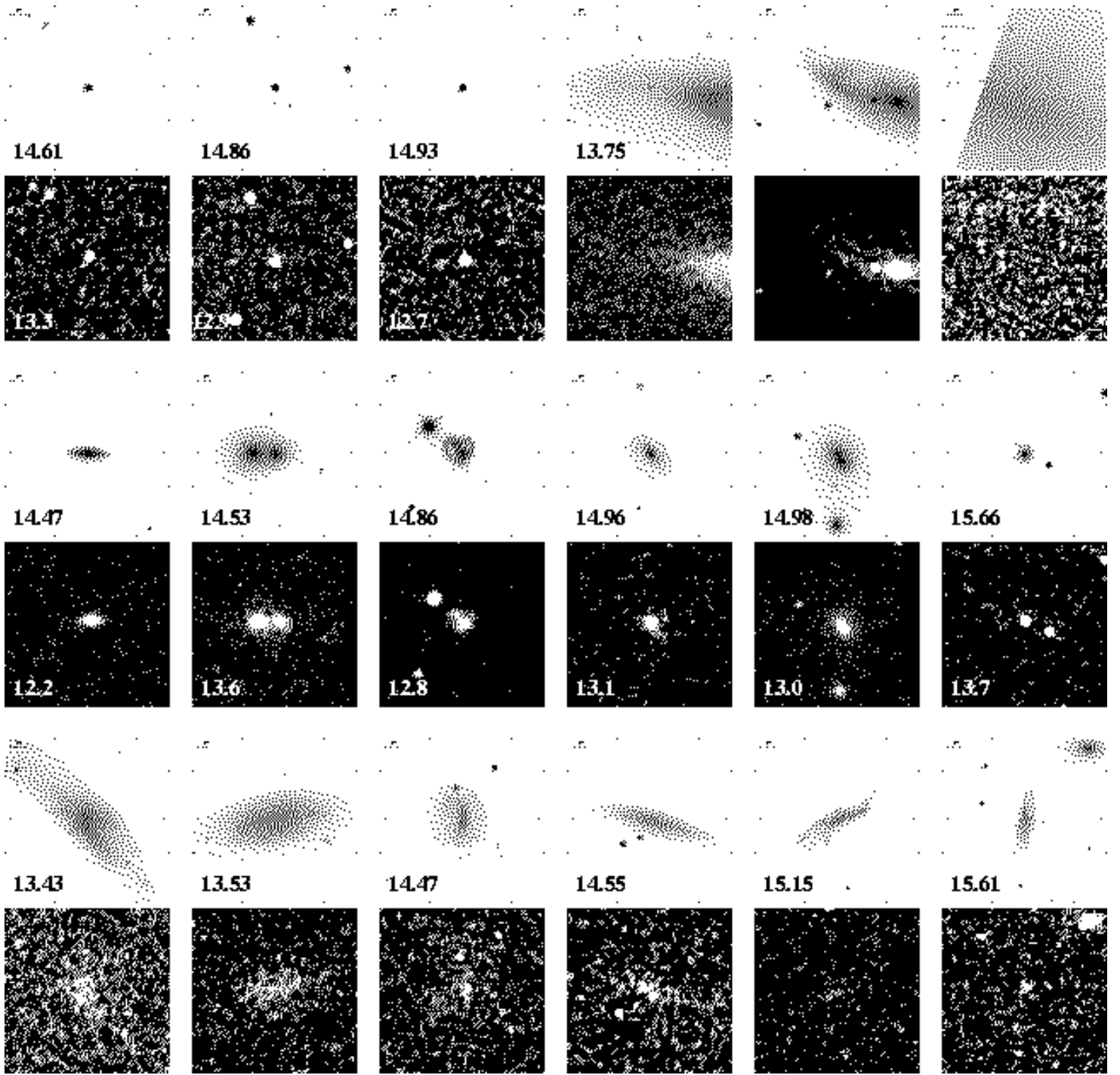}}
\caption{Examples of $r\leq16$ mag MGS objects that are not detected
in the XSC. For each row, we display the SDSS optical (top) and
2MASS $K$-band (bottom) images. We also show the extinction-corrected
Petrosian $r$ magnitudes and, for sources from the PSC, we indicate the
default aperture magnitude $K$, which is a rough measure of the
total galaxy light. Top row: the first three panels show sources
with PSC matches that appear
to be point-like (actual stars or unresolved nebulae), and the last
three panels give examples where SDSS falsely detects a bright galaxy as
the result of overlap from a large foreground galaxy, overdeblending of a
large galaxy, and scattered light from nearby bright star.
Second row: PSC-detected galaxies with bright round centres and
low-surface-brightness 
extended features; among these we show examples with nearby
bright companions from the XSC. Last row: typical examples of
late-type galaxies with low surface brightnesses that are below 2MASS
detection.
Images are $101\arcsec$ on a side, east is to the left, and a $10\arcsec$
line is given in the upper left of each optical image from SDSS.
\label{fig:various}}
\end{figure*}

We further investigate the low-level incompleteness at $r\leq16$ mag
with a visual inspection of the optical and near-IR image data at
the precise location of MGS objects without XSC matches.
First, we visually inspect all 232 (124 PSC, 108 unmatched) with $r\leq15$ mag.
We find that 42 (18\%) in fact have XSC counterparts,
but with 2MASS-SDSS coordinate offsets larger than the criteria we
use for cross correlation (see Appendix \ref{app:xcorrl}). We remind the
reader that our assessment of the completeness characteristics of
2MASS galaxies depends on our statistically-derived criteria that
assures negligible contamination from multiple and random matches.
Another 27\% (62) are false detections of bright galaxies
by SDSS. We find 40 are either faint background galaxies that have
falsely boosted flux as the result of overlap from the outer regions
of a large foreground galaxy or a very bright and nearby star, or are
part of a larger galaxy that was over deblended by the SDSS pipeline.
In addition, 22 of the PSC matches are actual point sources (stars or 
compact nebulae). We display examples of each of these false detections
in the top row of Figure \ref{fig:various}. The rest of the objects that
escape XSC detection are galaxies with visual characteristics
in agreement with what we find from Figure \ref{fig:all_cfs}.
Namely, bright galaxies in the PSC have bright round nuclei and
faint extended near-IR light that falls below the XSC detection, while
all unmatched galaxies are late-type systems that are too
low-surface-brightness to
be adequately detected by 2MASS (see examples in the bottom two rows of
Fig. \ref{fig:various}). We note that
$>1/3$ of the PSC galaxies have close companions of roughly equal
brightness. The companions are found in the XSC, thus, in these cases
the 2MASS processing must have attributed the common halo of each pair
to one galaxy (found in the XSC), and classified the second nucleus
as a point source. These galaxies have early-type morphologies, and
are possible ``dry merger'' candidates\footnote{Recent
studies \citep{vandokkum99,bell05b}
of $z>0.5$ early-type galaxies have found small numbers
of elliptical-elliptical galaxy mergers (so-called ``dry'' because
of the assumed lack of cold gas), which are likely an important route
for the formation of massive early-type galaxies.}.
If we remove the 62 false SDSS detections and include the 42 XSC matches
with large coordinate mismatches, we find an XSC completeness
of 97.4\% at $r\leq15$ mag.

Next, we inspect the images for a random 25\% sampling of the $15<r\leq16$
objects not found in the XSC (190 PSC, 117 unmatched). 
Even though our cross-correlation criteria
in this magnitude range is more restrictive, the fraction of examples that 
have XSC counterparts outside of our criteria is now less than 8\%.
We find 79 (26\%) with false SDSS galaxy detections, of which 62 are
the result of point sources making it into the MGS. The remaining
sources are LSBs (unmatched) and the bright cores of centrally-concentrated
galaxies, with a 3-fold decrease in the number of close pairs.
The bottom line here is that a complete visual inspection of over 1200
nonmatches could improve the already 92.4\% overall XSC completeness 
at $15<r\leq16$ mag by a few percent.

\subsection{Conditions for which the XSC Becomes Noticeably Incomplete}
\label{sec:incompl}

\begin{figure*}  
\center{\includegraphics[scale=0.88, angle=0]{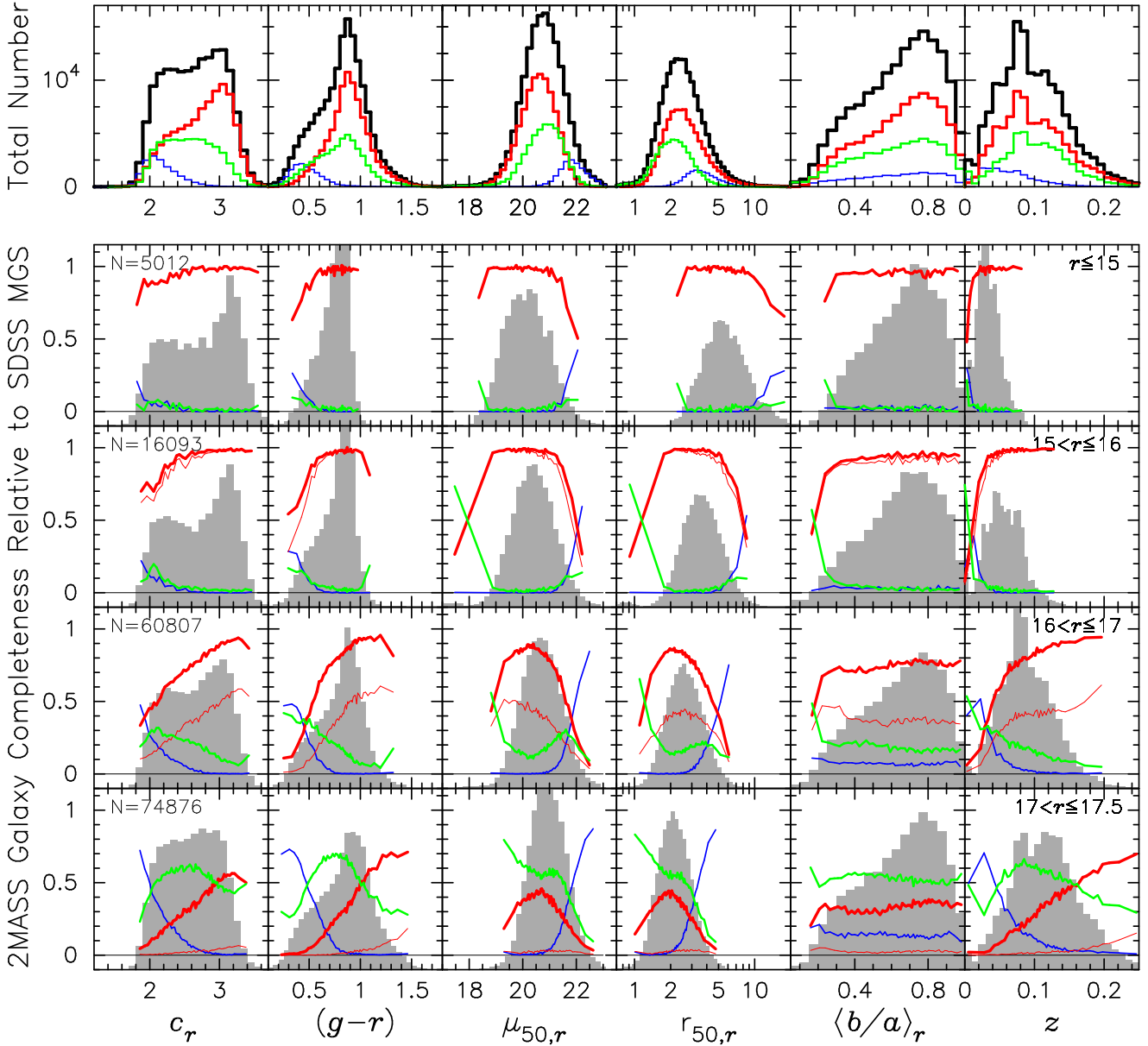}}
\caption[]{{\it Top row}: 
Observed optical property distributions for the SDSS ``main'' 
spectroscopic galaxy selection
(\mgssns; see text for details) limited 
to have extinction-corrected Petrosian magnitudes brighter
than $r=17.5$.  Shown from left to right are $r$-band concentration
given by $c_r={\rm r}_{90,r}/{\rm r}_{50,r}$, 
$(g-r)$ model colour, half-light $r$-band surface
brightness in mag/arcsec$^2$, Petrosian half-light radius in arcseconds
($r$-band), average $r$-band axis ratio $b/a$ (see \S \ref{sec:optnir}),
and spectroscopic redshift $z$. 
In each panel, we plot the total sample (black) of
156,788 galaxies with photometric and spectroscopic properties from DR2,
and the subsamples from our cross correlation with the 2MASS XSC 
(thick red, $N=89,742$) and PSC (green, $N=51,716$).
In addition, we show the subsample
of 15,330 galaxies without 2MASS counterparts (blue).
{\it Lower panels}: 
Completeness of 2MASS galaxies with respect to the SDSS spectroscopic
sample (\mgssns) as a function of six fundamental parameters. The completeness
functions (CF, see text for details) are given for four different $r$ magnitude
cuts corresponding to 95\%, 90\%, 50\%, and 10\% minimum XSC
completenesses per interval (see Fig. \ref{fig:rmag_cf}). 
We indicate the total number 
of galaxies per bin in the left-most panel.
The thin red line shows the $K\le13.57$ magnitude-limited XSC
sample, which contains a total of 43,663 galaxies. A thin black line marks
zero completeness in each panel, and for each magnitude interval 
we plot the \mgss property distributions in relative counts
(grey histograms).
\label{fig:all_cfs}}
\end{figure*}

Between Petrosian $r$-band magnitudes of 16.0 and 17.0,
the XSC galaxy completeness decreases rapidly from 90\% to only 50\%.
As shown in the fourth row of Figure \ref{fig:all_cfs},
we see many of the same trends as for brighter galaxies, i.e.
2MASS is most complete for more concentrated ($c_r>2.6$) and redder
($g-r>0.7$) galaxies with $\mu_{50,r}<21$~mag/arcsec$^2$.
In this $r$-magnitude interval, the differences in completeness
when including all sources from the XSC, and the subset brighter than
$K=13.57$ mag (thin red lines), becomes apparent. As one would
expect, a strict $K$-band magnitude cut omits galaxies that are
intrinsically fainter and bluer, which are typically lower in concentration
and surface brightness. The MGS sources not found in the
2MASS extended or point source catalogues concentrate towards the same
extremes of optical parameter space as at bright fluxes, that is 
blue, low-concentration, low-surface-brightness, and low redshift.

\begin{figure*}  
\center{\includegraphics[scale=0.99, angle=0]{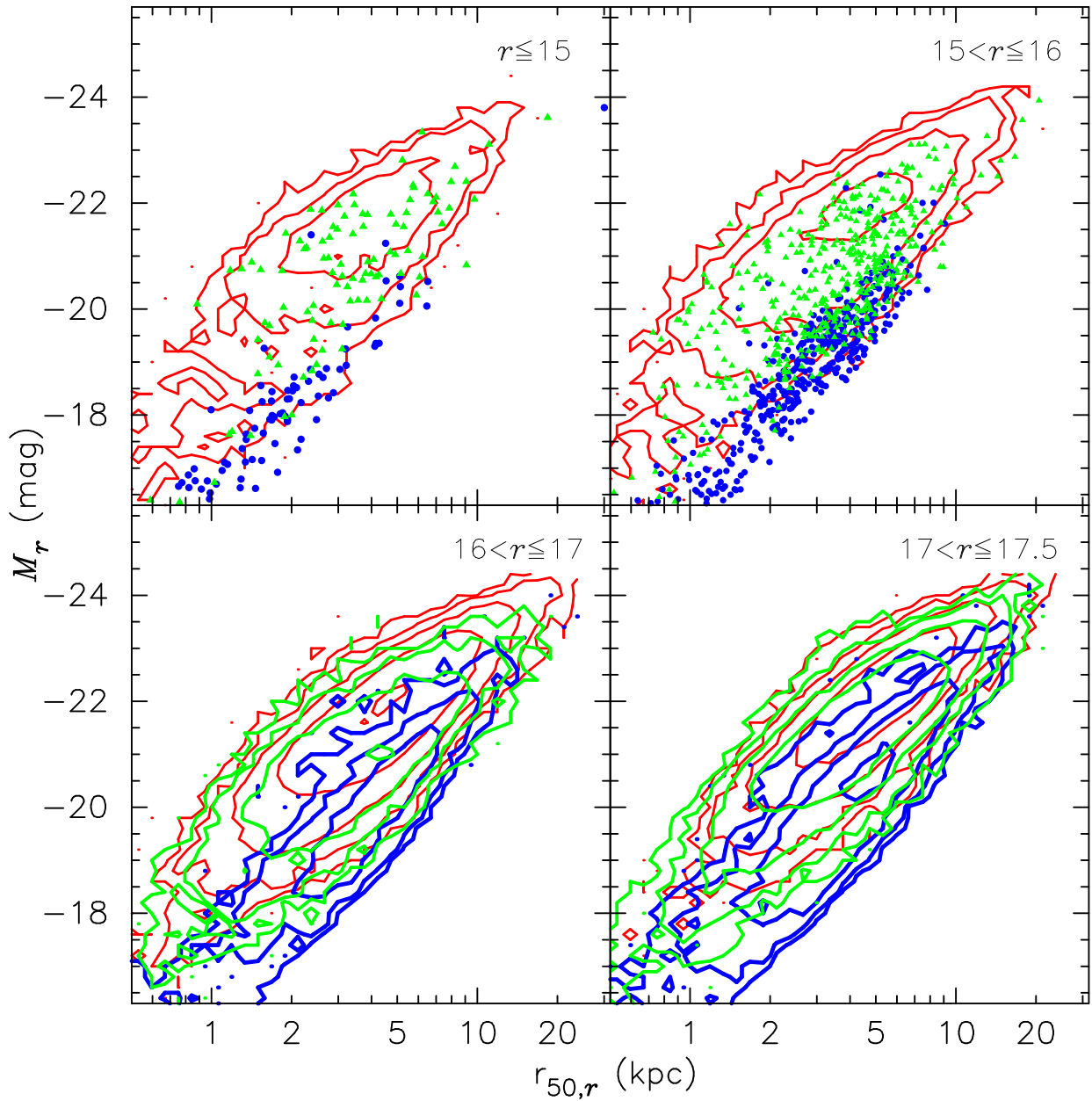}}
\caption[]{Petrosian $r$-band luminosity-size ($M_r-{\rm r}_{50,r}$) planes
for four $r$ magnitude intervals defined by minimum XSC completeness values
per bin of 95\%, 90\%, 50\%, and 10\% (see text for details).
In each panel we plot the absolute $r$-band magnitude against the
intrinsic half-light size in kiloparsecs. As in Fig. \ref{fig:all_cfs},
red denotes galaxies from the XSC, green the PSC, and blue represents
nondetections in 2MASS. Contours show levels of constant number density
in powers of three: 1, 3, 27, 81, 243, and 729.
\label{fig:rmagsize}}
\end{figure*}

\begin{figure}  
\center{\includegraphics[scale=0.85, angle=0]{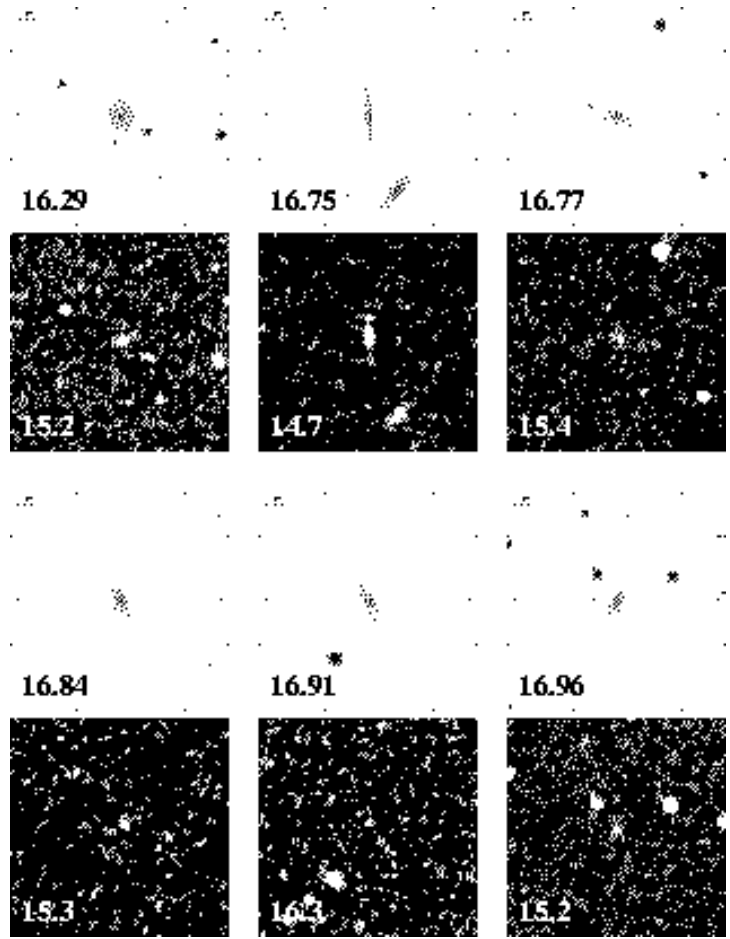}}
\caption{Examples of $16<r\leq17$ mag MGS galaxies with 
${\rm r}_{50,r}>3\arcsec$ that are detected as point sources by 2MASS.
As in Figure \ref{fig:various}, we display the optical and $K$-band
images for each example, and give the $r$ and $K$ brightnesses. 
Only the centres of faint disk-dominated galaxies in SDSS are detected
by 2MASS, which explains why they appear in the PSC.
\label{fig:faintPSCr3}}
\end{figure}

For $16<r\leq17$ mag, 
72\% of the MGS sources missed by XSC are present in the PSC, yet
a large fraction of these have $r$-band angular sizes that are big enough to
be resolved by 2MASS. Near the 2MASS resolution limit, we find 
that nearly $3/4$ (8188) of the PSC-detected galaxies
with $16<r\le17$ have ${\rm r}_{50,r}>2\arcsec$.
This subset of larger PSC galaxies have median parameter values of 
$(g-r)=0.58$, $c_r=2.36$, and $\mu_{50,r}=21.1$~mag/arcsec$^2$, 
which is suggestive of more late-type 
morphologies, and explains the changes in the behaviour of the 
parameter CFs compared with those at brighter $r$ cuts.
Indeed, a more conservative cut at ${\rm r}_{50,r}>3\arcsec$, which
is certainly resolved, still
comprises 35\% of the galaxies found in the PSC only.
In other words, we are seeing the limit at
which the disks and low-surface-brightness features of fainter
late-type galaxies fall below the detection limit of the XSC
algorithm, which in effect is primarily a cut in surface-brightness. 
To illustrate this point, in Figure \ref{fig:faintPSCr3} we
show SDSS and 2MASS images of several random PSC examples with
${\rm r}_{50,r}>3\arcsec$ and $16<r\leq17$ mag.
With the greater depth and resolution of SDSS we see
disk-dominated galaxies in the optical, but in the 2MASS images the
brighter cores of these sources appear as small, concentrated
objects and are point-like. At each successive $r$ cut we probe larger
redshifts on average, which results in more 
later-type galaxies falling below the detection limit of the XSC
algorithm. This is
confirmed by the $M_r-{\rm r}_{50,r}$ distribution of PSC detections
(lower left of Fig. \ref{fig:rmagsize}), where we see that the green
contours tend toward somewhat lower surface brightness
than the XSC galaxies; i.e., the PSC luminosity-size distribution is shifted
in the direction of lower average luminosity relative to that of the XSC.
The galaxies not found in 2MASS continue to populate
the low-surface-brightness part of this parameter space, with increased numbers
extending into the envelope defined by the XSC.

Finally, at the faintest half-magnitude interval of the complete SDSS
spectroscopic selection of galaxies (bottom panels of Fig. \ref{fig:all_cfs};
lower right of Fig. \ref{fig:rmagsize}),
we observe a continuation of the above trends
for the $16<r\le17$ interval. The surface brightness
limit of the XSC selection, combined with the fainter brightness and
greater redshift on average, extend the unmatched CFs further into 
each parameter distribution; therefore, we find that ever more concentrated,
redder, and high-surface brightness galaxies are missed altogether by 2MASS.
Moreover, more than half of
the MGS has only a point-source detection in 2MASS in this $r$ bin, with
51\% (8\%) of these having optical sizes of
${\rm r}_{50,r}>2\arcsec$ ($>3\arcsec$). Thus, most faint MGS galaxies
in the PSC are semi-resolved at best.
The PSC and XSC $M_r-{\rm r}_{50,r}$ relations are nearly identical,
with a growing number of nondetections populating a similar luminosity-size
space. The XSC is increasingly more complete for
the most distant galaxies in the MGS, which is expected due to their
typically redder colour.

\section{Nature of Galaxies Detected and Missed by 2MASS}
\label{sec:nature}

Now that we have quantified the detailed completeness of the 2MASS
galaxy sample, we move on to discuss further the nature of galaxies both
included in, and absent from, this $K$-band-selected sample. By
nature we mean inherent qualities such as rest-frame
luminosity, physical size, and colour. For this
we use the suite of SDSS and 2MASS measurements, corrected to rest-frame
$z=0$, to study the
optical and near-IR nature of galaxies in the XSC. Of particular interest
is the character of galaxies that 2MASS misses, determined by
quantifying the rest-frame properties of known galaxies
found only as 2MASS point sources or completely lacking detection in 2MASS.

\subsection{Optical Properties of Galaxies Outside of 2MASS Detection}
\label{sec:no}

Recall, the \mgss represents a representative sample of local galaxies
that is magnitude ($r\leq17.5$) and surface brightness 
($\mu_{50,r}\le 23.0$ mag/arcsec$^2$)
limited. Comparison with the median values of each property distribution
from the MGS (see Table \ref{tab:phot_spec_comp}) 
allows us to quantify which galaxies fall outside
of 2MASS detection. For example, among these galaxies, 95\% are less
concentrated than the median ($c_r=2.65$) MGS source.
The SDSS concentration has
been shown to correlate roughly with morphology \citep{shimasaku01,strateva01},
and a value of $c_r=2.6$ is commonly used to separate early and late
morphological types \citep{strateva01,kauffmann03,bell03a,bell03b}. This
value is very close to the MGS median; therefore, the galaxies
missed by 2MASS are systematically late type. As we
demonstrated in \S \ref{sec:gxcompl}, the average galaxy that escapes 2MASS
detection is much bluer and lower-surface brightness
than the typical MGS galaxy, fully consistent with late-type morphologies.
Indeed, even at optically bright magnitudes, these systems are visually
very late-type spirals, LSBs, or irregulars (see bottom row of
Fig. \ref{fig:various}).

As noted by \citet{strauss02},
the SDSS half-light sizes are not corrected for seeing, which biases
concentration values downwards for
galaxies with sizes comparable to the PSF size (typically $1\farcs4$).
The galaxies missed by 2MASS are larger
on the sky with a median size of $3.55\arcsec$
compared to the MGS median size of $2.46\arcsec$;
therefore, we are confident that the low-concentration nature of the
missed systems are not affected by seeing.
Furthermore, this difference in apparent size highlights why 2MASS
preferentially misses these galaxies. At a given $r$ magnitude, the
unmatched galaxies have the largest angular sizes, which translates
into the lowest apparent surface brightnesses. This is exactly
the behaviour one expects from a shallower magnitude-limited survey.
Moreover, the median redshift difference of $z_{\rm med}=0.057$
for nondetections, compared
to $z_{\rm med}=0.092$ for $r\le17.5$ mag galaxies from SDSS, owes
to this surface brightness selection effect.
This selection effect begins at $\mu_{50,r}\approx21.5$ mag/arcsec$^2$
(see Fig. \ref{fig:all_cfs}), or 1.5 mag brighter than the MGS limit.
Even at bright optical magnitudes where the XSC is $>90\%$ complete,
a fraction of the largest (${\rm r}_{50,r}>5\arcsec$) galaxies are
not detected. We see from Figure \ref{fig:nir.4rbins} that the XSC
detects galaxies larger than $5\arcsec$ but limited in mean surface
brightness to $\mu_{50,K}\leq19.5$ mag/arcsec$^2$, which is $0.5$ ($1$)
mag brighter (fainter) than the $1\sigma$ ($3\sigma$)
background noise in 2MASS $K$-band images \citep{jarrett00a}.
Therefore, we qualify the statement by
\citet{jarrett03} that galaxies larger than $10-15\arcsec$ in diameter,
but less than $50\arcsec$, are fully sampled in the XSC {\it if}
their $K$-band surface brightness averaged over the half-light diameter
is 0.5 mag above the 2MASS background. We note that the most
low-surface-brightness galaxies
in the 2MASS Large Galaxy Atlas are within this limit.

Finally, in Figure \ref{fig:optnature1} we
further demonstrate the different nature of MGS galaxies
not detected by 2MASS. Relative to the statistically-complete MGS
(shown in black), the galaxies
missed by 2MASS (blue histogram) are significantly lower in
luminosity and bluer, yet they have similar physical size distribution.
Therefore, the differences in
angular size between nondetections and all MGS galaxies are primarily
the result of their {\it intrinsic} low surface brightness,
i.e. at a given physical size, galaxies missed by 2MASS have lower
luminosity than typical galaxies detected by 2MASS.

\subsection{Optical Properties of Faint Galaxies in the PSC}
\label{sec:psc}

A large fraction (33\%) of MGS galaxies have 2MASS counterparts in
the PSC only; 98\% of these are fainter than $r=16$ mag
(see Table \ref{tab:rbin_totals}).
Galaxies have PSC-only detections mainly because they are
unresolved by 2MASS, with 44\% (85\%) having 
${\rm r}_{50,r}\leq2\arcsec$ ($3\arcsec$). The remainder have
larger angular sizes, but their surface-brightness profiles fall
below the 2MASS extended-source sensitivity and, thus, only their
round and concentrated nuclei are detected. As we show
in the middle row of Figure \ref{fig:various}, even at bright total
$r$ magnitudes the outer portions of these galaxies are faint in 
the optical, and nearly invisible in near-IR light. In terms of
morphology, the bright PSC detections that are large enough to be
resolved by 2MASS are either disk-dominated systems with 
low-surface-brightness
disks, or early-type galaxies in close pairs. At fainter than
$r=16$, these galaxies appear to be mostly disk-dominated (see
Fig. \ref{fig:faintPSCr3}).
We conclude that the late-type galaxies with higher-surface brightness
disks and large enough angular sizes
to be resolved by 2MASS
are detected readily in the XSC at $r\leq16$, while at fainter
magnitudes their disks fall systematically below detection.

For determining the morphological nature of PSC-selected galaxies 
with ${\rm r}_{50,r}\leq3\arcsec$, their small angular sizes
rule out comparisons with
the seeing-dependent concentration parameter from the overall sample.
Therefore, we turn to apparent $(g-r)$ model 
colour as a possible type discriminator. All galaxies in
the PSC have a median and r.m.s. colour of $(g-r)=0.81\pm0.25$,
which is well-matched to values for the MGS ($g-r=0.84\pm0.25$; see
Table \ref{tab:phot_spec_comp}). Yet, we find the resolved
and unresolved PSC galaxies have different median colours. The former
are moderately blue ($g-r=0.61$) as expected for 
later-type morphologies, while galaxies that are too small to be
resolved by 2MASS have $(g-r)=0.80$, which is close to the median
colour of MGS galaxies, suggesting a range of galaxy types.

In terms of rest-frame properties, we see in Figure \ref{fig:optnature1}
that the PSC detections (in green) span
similar relative distributions of optical luminosity, size,
and $(g-r)_0$ colour (corrected to $z=0$) as galaxies in the 
complete MGS (black). We notice
a slight bias such that the total PSC sample is somewhat bluer, 
less-luminous, and smaller, but this shift is explained by the small
fraction (15\%) of resolved galaxies with late-type morphologies and
low-surface-brightness disks that made it into the PSC. The
relative differences between the MGS and its PSC subsample are
minor for each property; therefore, we surmise that
truly unresolved galaxies detected by the PSC span the
full range of colours, luminosities, physical sizes, 
and presumably types, as normal galaxies. So it is simply the
larger cosmological distances ($z_{\rm med}=0.102$) that explain the
smaller angular sizes and
typically fainter galaxy content of the PSC.

\subsection{Optical and Near-IR Nature of Galaxies in the XSC}

\begin{figure*}  
\center{\includegraphics[scale=0.88, angle=0]{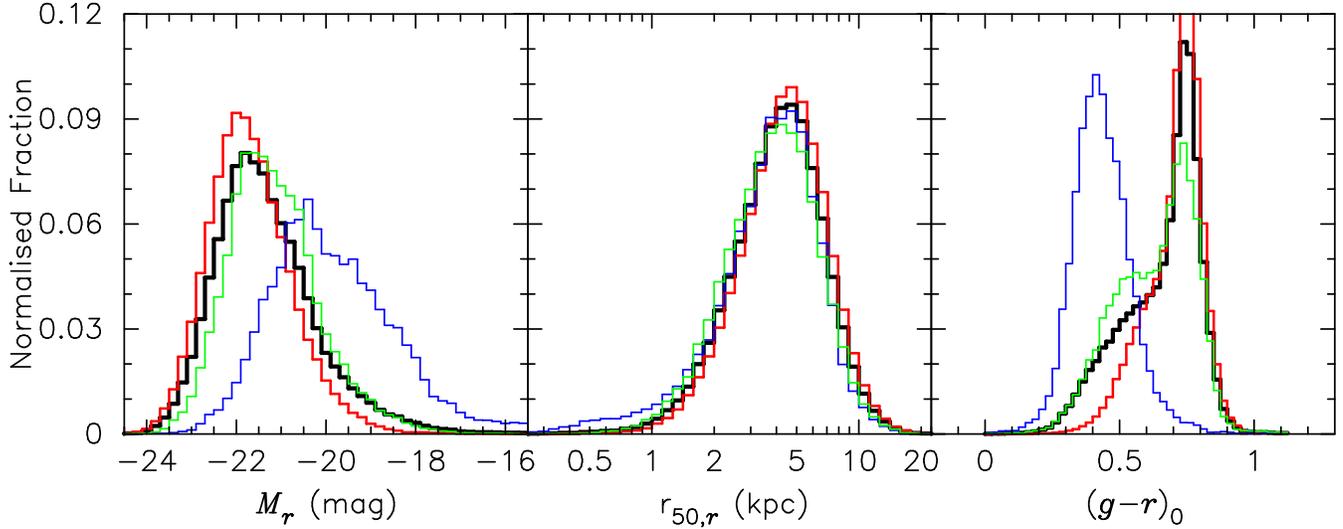}}
\caption{Normalised distributions of various 2MASS-related
selections of subsamples from the MGS ($r\le17.5$ mag), in terms of rest-frame 
optical properties:
absolute $r$-band magnitude (left), physical half-light radius in the $r$-band
(middle),
and rest-frame $(g-r)_0$ model colour (right). The magnitudes and colours
are $k$-corrected to $z=0$.
In each panel, we compare the overall distribution 
from the MGS (black) with distributions from those galaxies in the XSC (red),
in the PSC (green), and not detected by 2MASS (blue).
We truncate the XSC normalised $(g-r)_0$ distribution slightly at redder
colours so that we can resolve differences with other distributions
at bluer colours.
\label{fig:optnature1}}
\end{figure*}

We combine near-IR galaxy parameters from 2MASS with the SDSS
optical properties to describe the nature of 2MASS-selected galaxies.
We find that 57\% of $r\leq17.5$ galaxies from the MGS spectroscopic
sample have a match in the XSC, with the vast majority of nonmatches
fainter than $r=16$ mag. In Figure \ref{fig:optnature1}, 
we see small differences between the relative
distributions of all MGS galaxies and the subset matched to the XSC
(shown in red).
The XSC covers a very similar range in physical sizes, but we see a shift 
towards more high-luminosity and redder galaxies because the nondetections tend
to be intrinsically fainter and bluer as described in \S \ref{sec:no}. 

\begin{figure*}  
\center{\includegraphics[scale=0.88, angle=0]{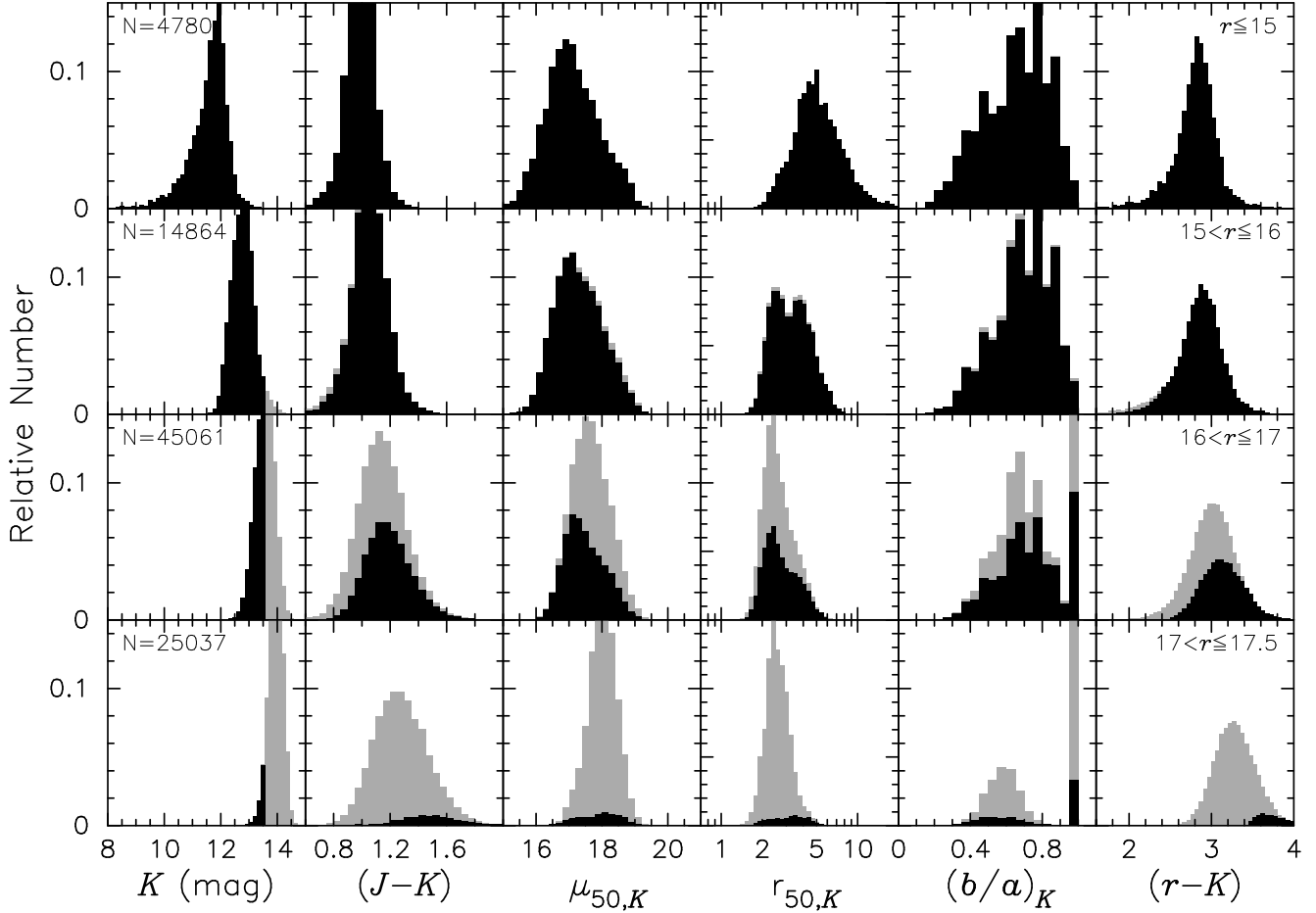}}
\caption[]{Near-infrared-derived property distributions from the XSC
for 2MASS galaxies in the SDSS spectroscopic sample (\mgssns) split
into four different $r$ magnitude intervals corresponding to
95\%, 90\%, 50\%, and 10\% minimum XSC completenesses per bin.
From left to right we give the apparent $K$-band Kron magnitude,
$(J-K)$ colour, average half-light surface brightness $\mu_{50,K}$,
circularised half-light radius ${\rm r}_{50,K}$, axis ratio $(b/a)_K$,
and Petrosian $r$ minus Kron $K$ colour.
We give the total distributions in grey and the $K\le13.57$ mag cut
in black. For each $r$ mag bin, we give the total number of XSC
galaxies in the left-most panel.
\label{fig:nir.4rbins}}
\end{figure*}

In Figure \ref{fig:nir.4rbins},
we plot the observed 2MASS parameter distributions for MGS galaxies
with XSC detections, split into four $r$ magnitude bins that 
represent decreasing XSC completeness with decreasing brightness. We see that
these galaxies span a broad range of apparent near-IR properties. For
optically bright galaxies ($r\leq16$ mag), the XSC is very complete
with nearly all sources having $K\leq13.57$ mag (black histograms). At
$r>16$ mag, many galaxies remain detected in the XSC, but with a 
growing number that have $K>13.57$ mag, 
which is the limit where 2MASS becomes nonuniform and
$<97.5\%$ complete. For $16<r\leq17$ mag, roughly half of the XSC
sources are fainter than $K=13.57$ mag, while at $r>17$ mag it is 92\%.
Naturally, this correspondence between optical
and near-IR faintness leads to the decreasing completeness of the XSC
at $r>16$ mag as we show in Figure \ref{fig:rmag_cf}.
Therefore, one must take care when defining
galaxy samples from cross-correlated 2MASS and SDSS observations.

\begin{figure*}  
\center{\includegraphics[scale=0.85, angle=0]{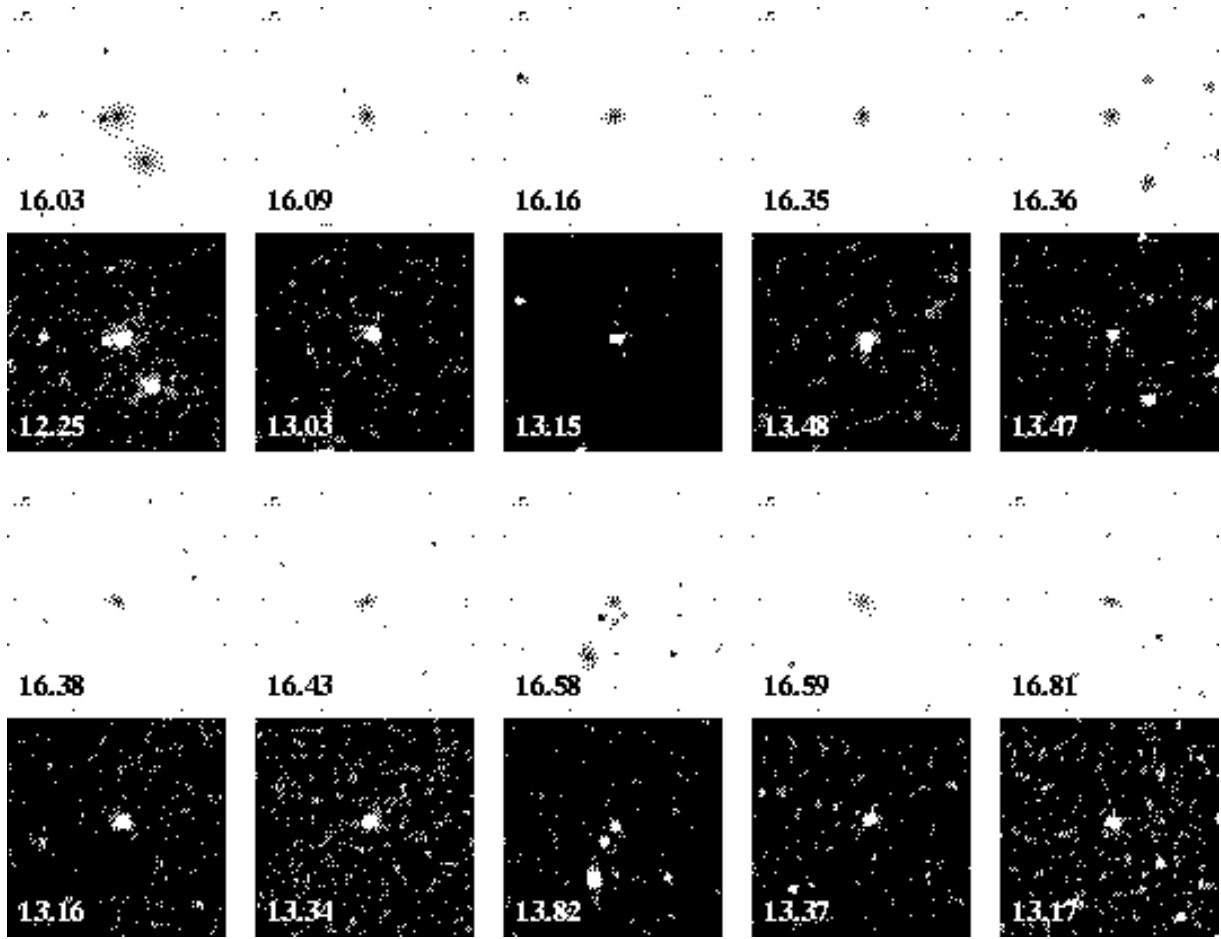}}
\caption{Examples of XSC galaxies with $16<r\leq17$ and red
colours ($1.0<g-r\leq1.2$). Following the format of Figures \ref{fig:noLSB},
\ref{fig:various} and \ref{fig:faintPSCr3},
we show $101\times101\arcsec$ optical and
$K$-band images for each case. Here we indicate the Petrosian $r$ and Kron
$K$ magnitudes. The galaxies are shown in decreasing $r$ brightness from
left to right. As illustrated here, galaxies meeting these criteria
are bulge-dominated 90\% of the time.
\label{fig:xscfaintred}}
\end{figure*}

For example, selecting all galaxies with $r\leq17.5$ and $K\leq13.57$
will provide a magnitude-limited sample that is 97.5\%
complete within the $K$-band limit, but includes $r>16$ galaxies that are
preferentially red and concentrated (see Fig. \ref{fig:all_cfs}). This colour
preference is also seen in the apparent $(J-K)$ and $(r-K)$ distributions of
$r>16$ galaxies in the bottom
two rows of Figure \ref{fig:nir.4rbins} (black filled histograms).
Here the $K$-band magnitude-limited XSC contains 62\% that is
redder than the median apparent $(g-r)$ colour of the MGS.
Clearly, the correlation between larger average redshifts for fainter
$r$ intervals produces overall shifts in the not-$k$-corrected
$(J-K)$ and $(r-K)$ colours towards
the red, which is likewise seen in the $(g-r)$ model colour
distributions of Figure \ref{fig:all_cfs}. Nevertheless, we have demonstrated
in \S \ref{sec:incompl} that the XSC preferentially selects from the
red side of the galaxy distribution at $r>16$ mag. 
In Figure \ref{fig:xscfaintred},
we show examples of galaxies selected at random from the region of
colour space ($1.0<g-r\leq1.2$) corresponding to the most complete XSC
galaxies at $16<r\leq17$ mag. These faint galaxies with 2MASS
detections are 90\% bulge-dominated, consistent with their red colours.

\begin{figure*}  
\center{\includegraphics[scale=0.85, angle=0]{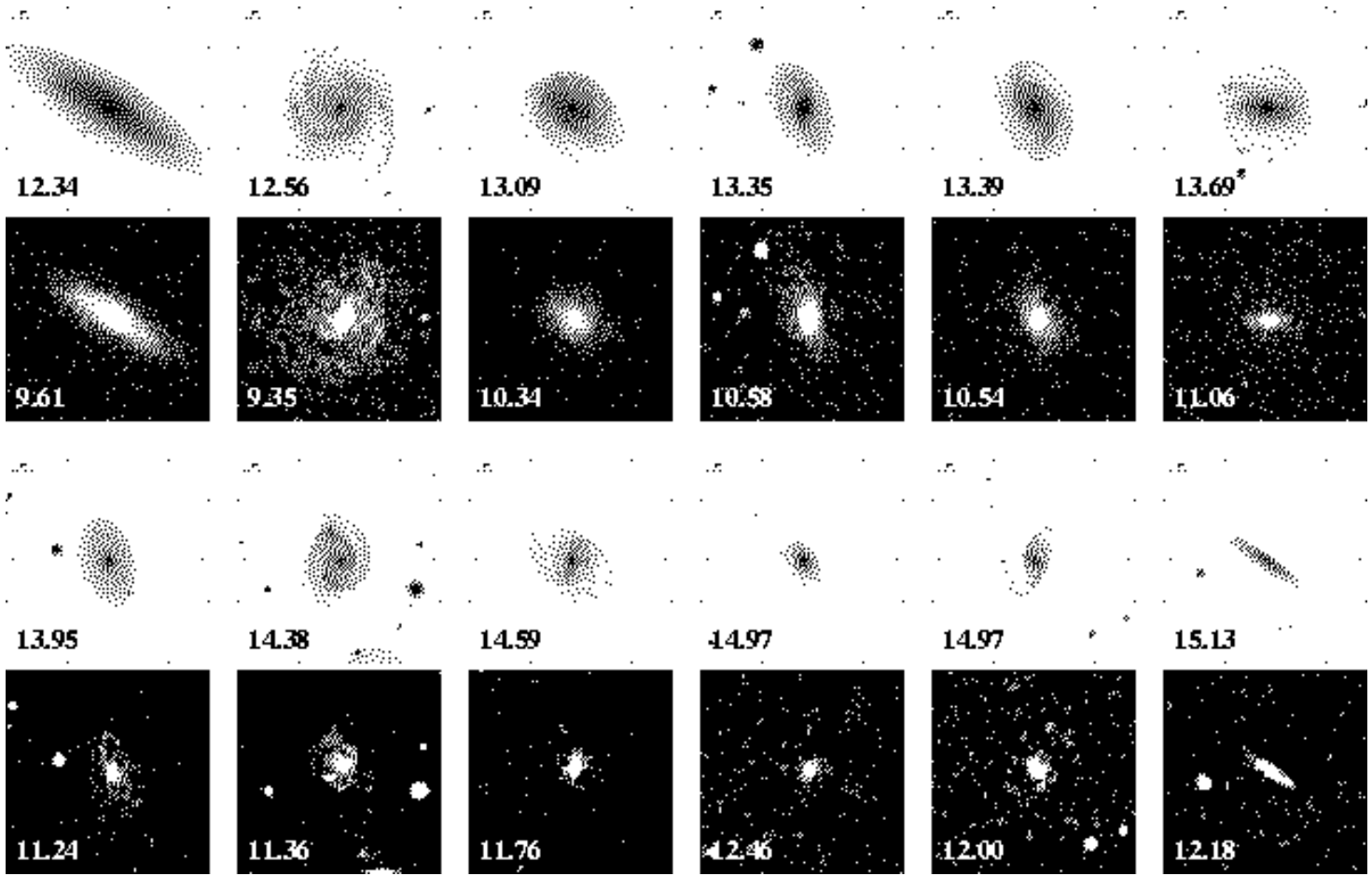}}
\caption{Examples of bright ($r\leq16$ mag) blue ($0.6<g-r\leq0.7$) 
galaxies found in the XSC. We display the images as in
Figure \ref{fig:xscfaintred}.
Roughly 90\% of galaxies with these $r$ and $(g-r)$
criteria are disk-dominated, and nearly $2/3$ of these have obvious
spiral structure in the SDSS images. We provide examples in these
relative frequencies.
\label{fig:xscblue}}
\end{figure*}

\begin{figure*}  
\center{\includegraphics[scale=0.85, angle=0]{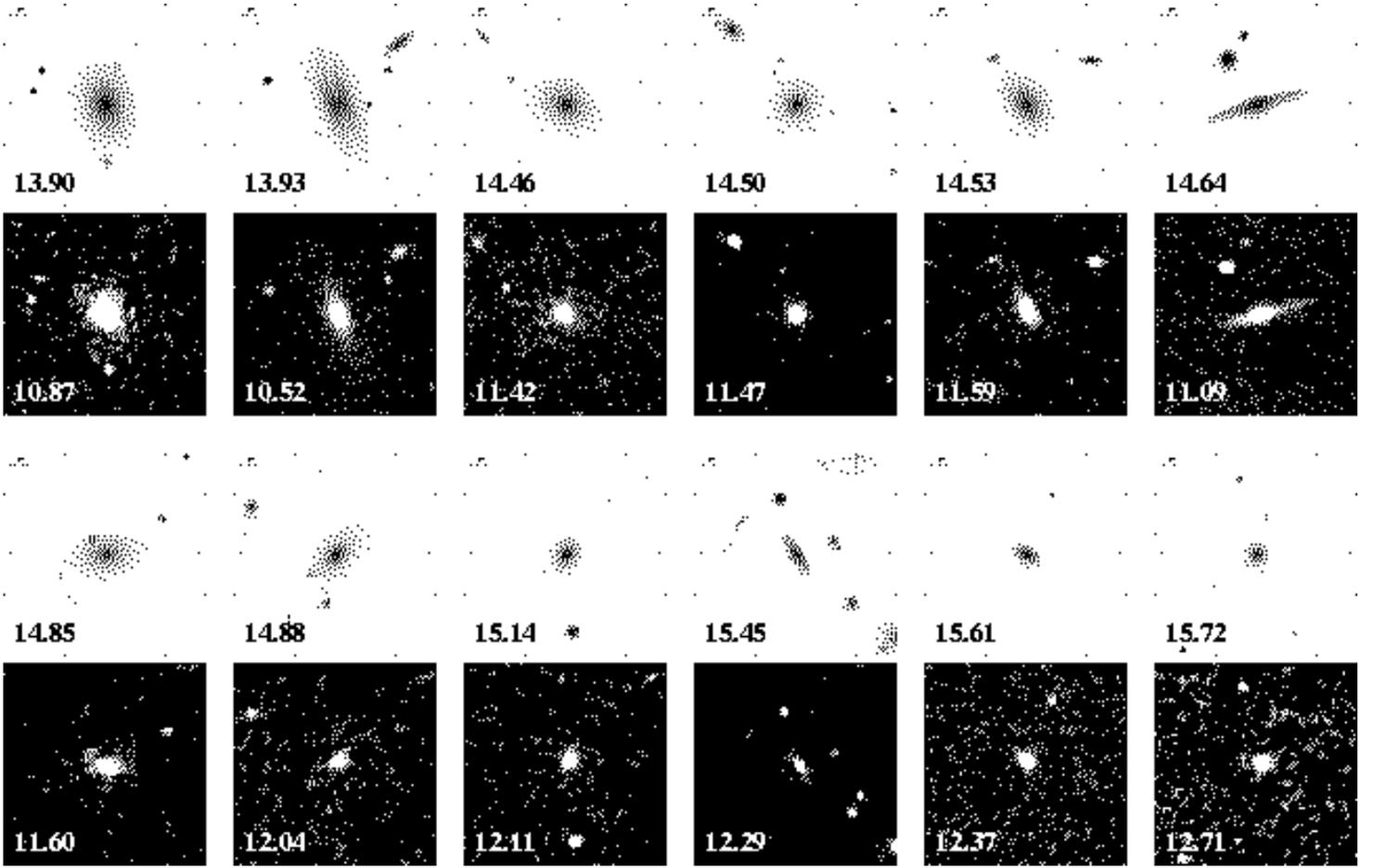}}
\caption{Same as in Figures \ref{fig:xscfaintred} and \ref{fig:xscblue},
here we show
examples of bright ($r\leq16$ mag) red ($0.9<g-r\leq1.0$)
galaxies with counterparts in the XSC. These galaxies
have bulge-dominated and disk-dominated frequencies of 85\%
and 15\%, respectively, as shown here.
\label{fig:xscred}}
\end{figure*}

Based on our analysis, selecting an XSC-MGS matched catalogue
that is magnitude-limited to $K\leq13.57$ and $r\leq16$ produces the most
representative sample with good statistical completeness (90.8\%)
and statistically useful numbers (19,156 for DR2;
hereafter, we call this sample XSC\_A). The near-IR cut ensures
97.5\% completeness of $K$-band galaxies in the XSC, and the $r$-band cut
provides 93.1\% completeness in terms of MGS
sources detected by the XSC.
The top two rows of Figure \ref{fig:nir.4rbins} illustrate clearly that
the XSC\_A is representative of the full range of near-IR parameter space,
and we have shown in \S \ref{sec:compl} that these galaxies are quite
representative in terms of observed optical properties. For reference,
we show examples of blue ($0.6<g-r\leq0.7$) and 
red ($0.9<g-r\leq1.0$) galaxies from the XSC with $r\leq16$ mag
in Figures \ref{fig:xscblue} and \ref{fig:xscred}. We choose these colour cuts
on either side of the median MGS colour such that the XSC selection is
$>95\%$ complete for sources within this range of colours.
Approximately 90\% of the blue galaxies are disk-dominated,
the majority with obvious spiral features, while the red galaxies have
early-type (bulge-dominated) morphologies. These figures demonstrate
the normal nature of galaxies detected with 93.1\% completeness in 2MASS.

\begin{figure*}  
\center{\includegraphics[scale=0.88, angle=0]{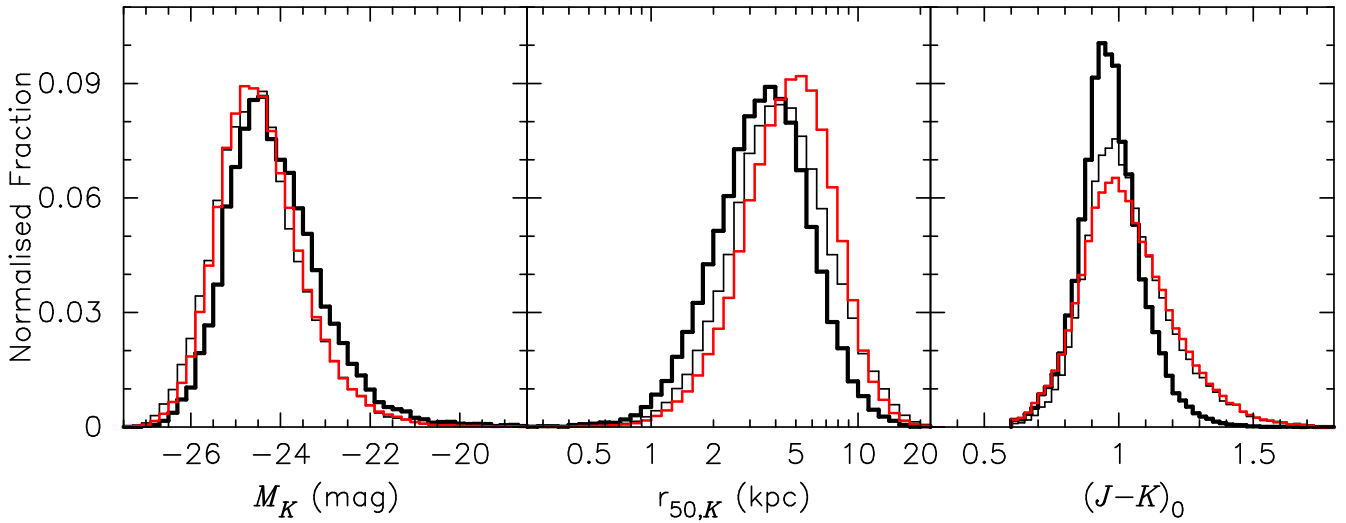}}
\caption{Relative normalised distributions of different $K$-magnitude
cuts to the XSC-MGS matched sample in terms of rest-frame near-IR properties:
absolute $K$-band magnitude (left), physical half-light radius in $K$-band
(middle),
and $(J-K)_0$ colour (right). The magnitudes and colours
are $k$-corrected to $z=0$.
In each panel, we compare the distribution 
from all XSC-MGS galaxies (red; XSC\_C) with distributions from subsets cut
at $K\le13.57$ mag (thin black; XSC\_B), and the most complete sample
limited at both $K\le13.57$ and $r\le16$ mag (thick black; XSC\_A).
\label{fig:nirnature}}
\end{figure*}

\begin{figure*}  
\center{\includegraphics[scale=0.88, angle=0]{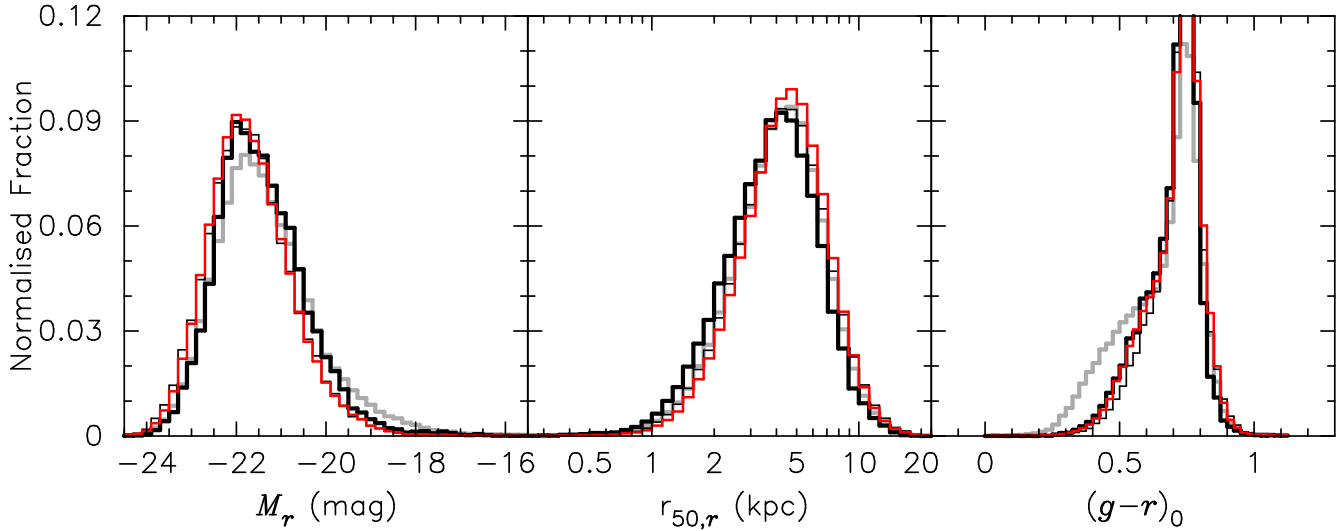}}
\caption{Relative normalised distributions of rest-frame optical properties
as in Figure \ref{fig:optnature1}, but here for the different $K$-magnitude
cut subsamples of the XSC-MGS matches as in Figure \ref{fig:nirnature}:
XSC\_A (thick black), XSC\_B (thin black), and XSC\_C (red).
We include the overall distribution from the MGS in grey for reference.
We truncate the XSC\_C normalised $(g-r)_0$ distribution slightly at redder
colours so that we can resolve differences with other distributions
at bluer colours.
\label{fig:optnature2}}
\end{figure*}

We compare the intrinsic nature (luminosity, size, colour) of XSC\_A
with two other samples based on the XSC-MGS cross correlation: 
the $r\leq17.5$ MGS matched
to the XSC (XSC\_B), and the subset limited in $K$-band to brighter
than 13.57 mag (XSC\_C). Using the DR2-derived MGS, XSC\_B and XSC\_C
contain 89742 and 43663 galaxies, respectively. We plot the
relative property distributions at near-IR wavelengths in
Figure \ref{fig:nirnature}, and at optical in Figure
\ref{fig:optnature2}. Not surprisingly, applying a strict $r$ magnitude
cut to a $K$-band-limited sample (XSC\_B $\rightarrow$ XSC\_A) results in an
additional blue-colour selection, which is well-illustrated by the
$(J-K)_0$ differences in the right panel of Figure \ref{fig:nirnature}.
Furthermore, the $r=16$ mag limit restricts the volume of space from
which these samples are drawn. Therefore, the XSC\_A sample contains
fewer luminous ($M_K<-24.5$ and $M_r<-22.0$) and large ($>4$ kpc)
galaxies compared to the XSC\_B and XSC\_C. The smaller median redshift
of XSC\_A ($z_{\rm med}=0.052$) compared to the XSC\_B (0.115) and
XSC\_C (0.093) confirms these trends.

\section{Conclusions} 
\label{sec:conc}

In this paper we have addressed the basic question: What is the nature
of galaxies detected and missed by 2MASS? 
We have matched 2MASS extended and point sources to the
spectroscopically-confirmed sample of over 155,000
bright ($r\leq17.5$ mag) galaxies from the SDSS (MGS).
We assume that the MGS provides a globally
representative sample of local galaxies with which to quantify
the average optical properties of galaxies that 2MASS does and does not detect.
From the cross correlation with MGS we find that 2MASS detects 90\%
of SDSS galaxies brighter than $r=17$ mag. We find that naively applying
the MGS criteria to a photometric-only selection from SDSS produces a less
reliable set of galaxies with a $10-20\%$ contamination by image artifacts 
in the SDSS at $r<15$ mag.

The 2MASS Extended Source Catalog (XSC) 
detects 93.1\% of all galaxies in the spectroscopic MGS down to $r=16$ mag.
If one does a careful job of matching to MGS and weeding out spurious SDSS
sources, a completeness of $97.5\%$ can be achieved at $r\leq15$ mag,
with the only major source of incompleteness being blue LSBs.
We have quantified the completeness in terms of
fundamental optical properties from SDSS -- concentration, colour, mean 
surface brightness, half-light radius, axis ratio, and redshift. We
find that the bright galaxies in the XSC span the full range of
optical properties exhibited by the SDSS main sample.
The XSC selection of known galaxies suffers two primary systematics:
(1) it is limited at the low-surface-brightness end as expected
in a magnitude-limited selection; and (2) it has a spatial resolution
limit of $2-3\arcsec$, which results in galaxies with smaller angular
sizes being selected and placed in the 2MASS Point Source Catalog (PSC).
The latter effect occurs mainly for galaxies fainter than $r=16$ mag,
which tend to be further away and, hence, smaller in angular extent.
Fewer than 3\% of $r\leq16$ galaxies from the MGS escape 2MASS detection 
because they have low-surface-brightness
and late morphological type
(e.g., Sc-Sd spirals, irregulars).
Another 4\% are detected only by the PSC as a result of their bright
round centres and faint outer profiles. Therefore, we conclude that the
low-level incompleteness in the XSC at the bright end is caused by
surface brightness selection effects.

At $16<r\leq17$, where the MGS remains highly complete, 2MASS continues
to detect galaxies at 90\% down to $r=17$ mag but with increasing
proportion in the PSC only. The majority (85\%) of
faint PSC detections are unresolved ($<3\arcsec$)
by 2MASS, with a broad range
of normal properties and redshifts towards the more-distant tail of the
distribution. As a result of the surface brightness-dependent selection
effects, 15\% of faint PSC galaxies are resolved disk-dominated systems
with disks too faint to detect in 2MASS. Finally, 2MASS still finds more than
80\% of the MGS fainter than $r=17$ mag, with more than half in the PSC
and only the most distant, red,
and presumably early-type galaxies found in the XSC.

Because of the different selection effects of 2MASS and SDSS, one must
be careful when defining samples of XSC-MGS matched galaxies based on
statistically-complete magnitude cuts from each individual survey. For example,
selecting all MGS galaxies with XSC matches will produce a sample that
is incomplete at $r>16$ mag and over-representative of red early-type
galaxies. Likewise, all $K\leq13.57$ galaxies from the XSC are complete
in a $K$-magnitude sense, but preferentially miss blue and lower-surface
brightness disk-dominated galaxies at $r>16$ mag. We note that all MGS
galaxies brighter than $K=13.57$ are fully representative of the half
million galaxies of similar brightness in the local universe as seen by 2MASS.
This means that 2MASS {\it does not} find some special type of galaxy not
present in the SDSS spectroscopic selection. Finally,
we find that a combined $K\leq13.57$ and $r\leq16$ cut produces the most
representative sample. Based on the DR2 version of the MGS
selection, this sample contains 19,156 galaxies with 90.8\%
completeness, spectroscopic redshifts, and a suite of near-IR
and optical properties.

\section*{Acknowledgements}
We are very grateful to Rae Stiening, who has tirelessly maintained
the {\it Two Micron All Sky Survey} (2MASS) catalogues and image data
at the University of Massachusetts, and who has helped write
efficient algorithms for searching the data base.
We thank Roc Cutri, Tom Jarrett, and Steve Schneider
for help with 2MASS.
We are grateful for enlightening correspondence with 
Sloan team members Jennifer Adelman, Michael Blanton, Sebastian Jester, Huan Lin, 
Michael Strauss, and Douglas Tucker.  
Additionally, the authors appreciate useful discussions with Rose Finn,
Kelly Holley-Bockelmann, and Ariyeh Maller.
D.\ H.\ M.\, M.\ D.\ W.\, and N.\ K.\ 
acknowledge support from the National Aeronautics
and Space Administration (NASA) under LTSA Grant NAG5-13102 
issued through the Office of Space Science.
M.\ D.\ W.\ and N.\ K.\ 
acknowledge support by NSF grants AST-9988146 \& AST-0205969 and
by the NASA ATP grant NAG5-12038.
E.\ F.\ B.\ was supported by the European Community's Human Potential
Program under contract HPRN-CT-2002-00316 (SISCO).
This publication makes use of data products from 2MASS
and the {\it Sloan Digital Sky Survey} (SDSS). 2MASS is a joint project of the
University of Massachusetts and the Infrared Processing and 
Analysis Center/California Institute of Technology, funded by 
the National Aeronautics and Space Administration and 
the National Science Foundation. The 2MASS Web site is
{\texttt http://pegasus.astro.umass.edu/}.
We obtained SDSS data from the SDSS Catalog Archive Server
({\texttt http://casjobs.sdss.org/casjobs/}), and
colour images presented here were downloaded with the SkyServer Image List
Tool ({\texttt http://cas.sdss.org/astro/en/tools/chart/list.asp}).
Funding for the creation and distribution of the SDSS Archive
has been provided by the Alfred P.\ Sloan Foundation, the 
Participating Institutions, the National Aeronautics and Space Administration,
the National Science Foundation, the U.S. Department of Energy, 
the Japanese Monbukagakusho, and the Max Planck Society.  The SDSS
Web site is {\texttt http://www.sdss.org/}.  The SDSS is managed by the
Astrophysical Research Consortium (ARC) for the Participating Institutions,
which are the University of Chicago, Fermilab, the Institute
for Advanced Study, the Japan Participation Group, the Johns Hopkins
University, Los Alamos National Laboratory, 
the Max-Planck-Institute for Astronomy (MPIA), the Max-Planck-Institute for
Astrophysics (MPA), New Mexico State University, University of Pittsburgh,
Princeton University, the United States Naval Observatory, and 
the University of Washington. 
This publication also made use of NASA's Astrophysics Data System 
Bibliographic Services.

\bibliographystyle{mn2e}
\bibliography{/home/dmac/Papers/refs.4mnras}

\appendix
\section{Defining Optimum Criteria for Matching SDSS Main Galaxies With 2MASS XSC Counterparts}
\label{app:xcorrl}
For every MGS photometric source from DR2 we first
find the nearest XSC source within a $60\arcsec$ radius. This allows
us to characterise the criteria
that we will use to construct the final
XSC-MGS matched catalog of galaxies.
Obviously, at larger angular separation $\theta_{\rm sep}$ between
DR2 and XSC source centres there is an increased chance of multiple and 
random matches.
The degree of this contamination in XSC-MGS cross correlation
depends on magnitude and $\theta_{\rm sep}$. Since all galaxies
follow a magnitude-size relation, such that brighter galaxies are
larger on average, there is greater chance for a larger
angular separation between the centres of DR2 and XSC
cross-correlated sources.

\begin{figure*}  
\center{\includegraphics[scale=1, angle=0]{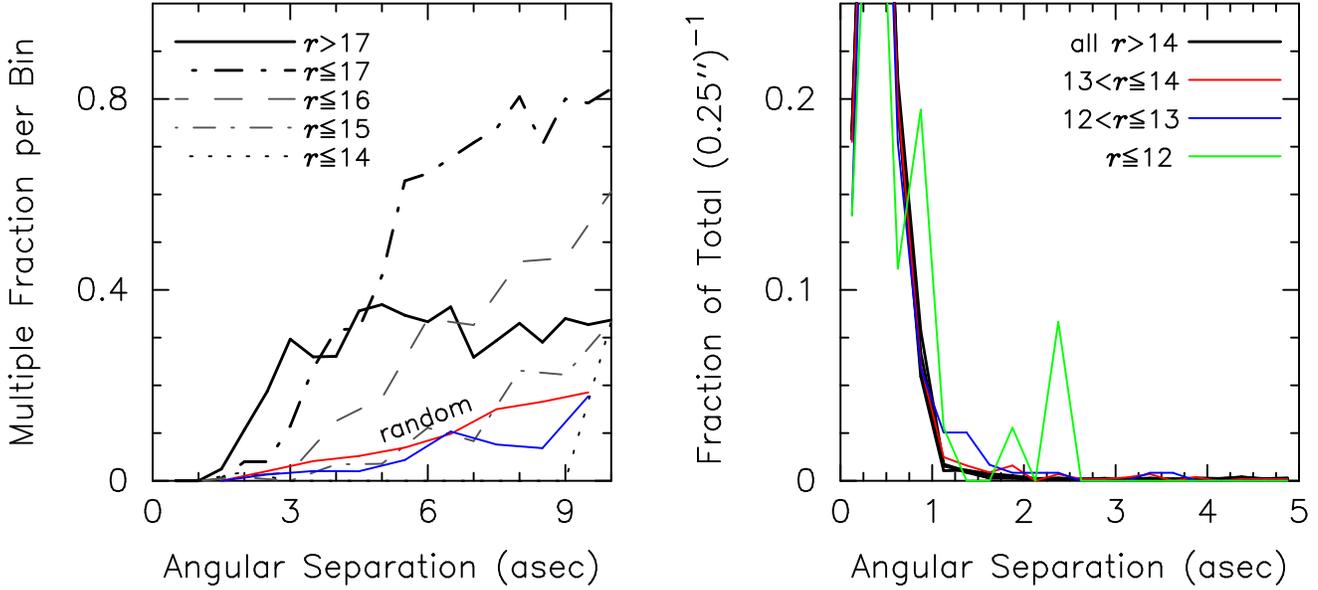}}
\caption{
Multiple 2MASS XSC sources matched to individual SDSS MGS sources. 
Left: Fraction of multiple sources per angular separation bin
for different $r$-band magnitude cuts (not extinction corrected).
The $\theta_{\rm sep}$ bin size is $1\farcs0$ for $r\leq 16$ mag 
(bold lines) and
$0\farcs5$ for $r>16$ mag (thin lines). All sources have $\mu_{50,r}\leq23$.
In red ($16<r\leq17$) and blue ($14<r\leq15$), we estimate the number of
random matches for the two $r$ mag intervals.
Right: For each $r$-band magnitude bin,
we calculate the fraction of
matched sources per $0\farcs25$ interval relative to the
total number of cross-correlated sources within a $5\arcsec$
radius.
\label{fig:xcor_info}}
\end{figure*}

We consider a multiple cross correlation to be two (or more) DR2 sources 
with the
same XSC match. In the left panel of Figure \ref{fig:xcor_info}, we 
plot the fraction of multiple sources per 
$\theta_{\rm sep}$ bin for different $r$-band magnitude limits. 
The multiple fraction increases more rapidly and at smaller angular separation
for fainter magnitude cuts as a result of the increasing DR2 number counts.
This behaviour continues
until the final cut of $r>17$ mag, at which point the XSC sources become so
incomplete that the chance of even a random match at larger $\theta_{\rm sep}$
diminishes, making the observed multiple fraction flatten. Moreover,
we illustrate the estimated fraction of random chance cross correlations
per separation bin. The number of random sources $N_{\rm rand}$
is given by the
probability of finding a XSC source in a given search area
($\pi\theta_{\rm sep}^2$) multiplied by the number of DR2 sources within
that area. The
probability of finding a random 2MASS XSC source in one square arcsecond is
$3.58e-6$, given by 956,221 total sources over one half of the sky
($|b| \ge 30\arcdeg$). For
example, there are 60,601 SDSS $16<r\leq17$ ($\mu_{50,r}\leq23$)
sources with $\theta_{\rm sep}\leq3\farcs0$
resulting in 6 random matches, while there are only an additional 116 SDSS 
matches if
we increase to $\theta_{\rm sep}\leq4\farcs0$ and an additional
5 random matches or 4\% random fraction for this bin. Similarly,
there is roughly 2\% random chance in the same interval for SDSS $14<r\leq15$
sources.
We find that the random fraction contamination slowly
diminishes at brighter magnitudes as a result of the lowering DR2 number
counts.

In addition, we examine the distribution of nearest angular separation
matches in the right panel of Figure \ref{fig:xcor_info}. Here we plot the
fraction of sources per $0\farcs25$ separation bin relative to the total 
number of sources with $\theta_{\rm sep}\leq5\farcs0$ for
different $r$-band magnitude intervals. All but the two brightest ($r<13$ mag)
intervals have identical angular separation distributions that peak around
$\theta_{\rm sep}=0\farcs5$, then fall quickly to $<1$ percent at 
$\leq1\farcs25$, and finally slowly drop to zero around $2\farcs0$. We see
that a non-negligible fraction of the $r<13$ mag sources have separations
$>1\farcs25$.

By combining the information in both panels of Figure \ref{fig:xcor_info}
we can define an upper limit on the
maximum angular separation per $r$-band magnitude interval that
maximises the number of useful XSC-MGS cross-correlated sources
and maximises contamination from multiple and random matches.
For example, the left panel shows that multiple and random fractions for 
$r\leq17$ and $r\leq16$ are $<1\%$ for $\theta_{\rm sep}$
of $1\farcs5$ and $2\farcs0$, respectively. For brighter magnitudes we
see that larger separations are allowed before the contamination
from multiple sources exceeds 1 percent.
In columns 1--3 of Table \ref{tab:xcor}, we tabulate the optimised
angular separation cut per $r$-band magnitude interval, and the
total number $N_{\rm cut}$ of matched sources therein. For the two
bins brighter than $r=13$ mag, we determine the best $\theta_{\rm sep}$
criteria by visual inspection of all matches within $60\arcsec$ radius.

In Table \ref{tab:xcor}, we include the total counts from matching within
other $\theta_{\rm sep}$ cuts ($5\farcs0$, $2\farcs0$, and $1\farcs25$),
and an estimate of the number of multiple matches for several
separations ($2\farcs0$, $3\farcs0$, and $5\farcs0$). For the magnitude bins
brighter than $r=15$, we visually confirm the
optimally-matched sources with $\theta_{\rm sep}>1\farcs25$. Recall
that nearly all matches have separations less than $1\farcs25$
(Fig. \ref{fig:xcor_info}, right panel). The number
of sources between $1\farcs25$ and our optimised cut is given in column 10
of Table \ref{tab:xcor}, and is the difference between columns 3 and 6.
For $r\leq15$ mag, this number is $>1\%$ of $N_{\rm cut}$, and
thus, warrants visual inspection.

In summary, we define a sliding angular separation cut that depends
on raw $r$-band magnitude to achieve the most thorough catalog of XSC-MGS
cross-correlated sources with negligible contamination from multiple and
random matches. These criterion account for the extended nature of galaxies
in the XSC. For matching SDSS main galaxies with 2MASS point sources, we
search the PSC only for MGS sources not found in the XSC using a simple
nearest match within a $2\arcsec$ radius (see \S \ref{sec:xcorrl}).

\begin{table*}
\caption{XSC-MGS cross correlation.}
\label{tab:xcor}
\begin{tabular}{cccccccccc}
\hline\hline
 mag bin & $\theta_{\rm sep}$ cut & $N_{\rm cut}$ & $N$ & $N$ & $N$ & $N_{\rm mult}$ & $N_{\rm mult}$ & $N_{\rm mult}$ & $N$ \\
   & & & $(\leq5\farcs0)$ & $(\leq2\farcs0)$ & $(\leq1\farcs25)$ & $(\leq2\farcs0)$ & $(\leq3\farcs0)$ & $(\leq5\farcs0)$ & $(1\farcs25,{\rm cut})$ \\
(1) & (2) & (3) & (4) & (5) & (6) & (7) & (8) & (9) & (10) \\
\hline
$r\leq12$     & $30\arcsec$ &    40 &    36 &    33 &    32 &  0 &  0 &   0 & 8\\
$12<r\leq13$ & $10\arcsec$ &   242 &   237 &   233 &   224 &  0 &  0 &   0 & 18\\
$13<r\leq14$ & $4\arcsec$  &  1124 &  1127 &  1110 &  1087 &  1 &  1 &   1 & 37\\
$14<r\leq15$ & $3\arcsec$  &  4853 &  4897 &  4832 &  4774 &  0 &  0 &   1 & 79\\
$15<r\leq16$ & $2\arcsec$  & 19149 & 19335 & 19149 & 18959 &  4 &  6 &  12 & 190\\
$16<r\leq17$ & $1\farcs5$  & 60267 & 60836 & 60432 & 59965 &  6 & 19 &  86 & 302\\
$r>17$       & $1\farcs25$ & 56106 & 57373 & 56548 & 56106 & 39 & 95 & 283 & 0\\
\hline
\end{tabular}
\vskip 12pt
\begin{minipage}{\hdsize}
For the optimised XSC-MGS cross correlation per raw
Petrosian $r$-magnitude intervals (1), we give the 
angular separation $\theta_{\rm sep}$ criteria between MGS and XSC source 
centres (2), and the resulting number of matched sources (3).
The total number of matched sources
using this set of criteria is 141,781 (sum of column 3). We provide total
(4--6) and multiple (7--9) source counts for additional $\theta_{\rm sep}$
cuts. We visually inspect all optimally-matched sources (10)
with $r\leq15$ mag and $\theta_{\rm sep}>1\farcs25$.
\end{minipage}
\end{table*}

\bsp

\label{lastpage}

\end{document}